# EarthFinder Probe Mission Concept Study

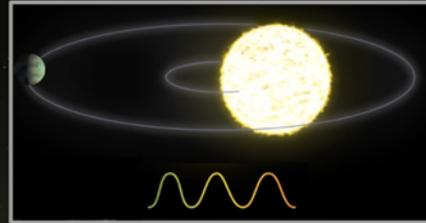

Characterizing nearby stellar exoplanet systems with Earth-mass analogs for future direct imaging

March 2019

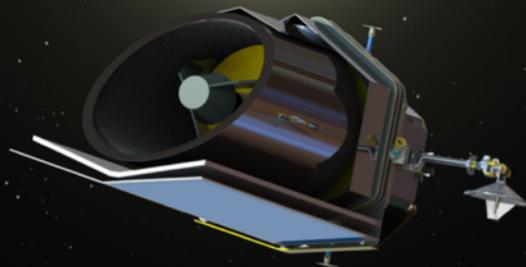

NASA Aeronautics and Space Administration

PI: Peter Plavchan
George Mason University
Fairfax, Virginia

Co-I: Gautam Vasisht
Jet Propulsion Laboratory
California Institute of Technology
Pasadena, California

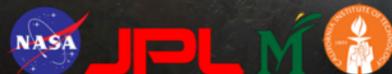

## Team Members

*Peter Plavchan (PI, George Mason University)*

*Gautam Vasisht (Co-I, Instrument team lead, NASA Jet Propulsion Lab)*

*Chas Beichman (Instrument team lead, NASA Exoplanet Science Institute)*

*Heather Cegla (Stellar Activity team lead, U. Geneva)*

*Xavier Dumusque (Stellar Activity team lead, U Geneva)*

*Sharon Wang (Telluric team lead, Carnegie DTM)*

*Peter Gao (Ancillary Science team lead, UC Berkeley)*

*Courtney Dressing (Ancillary Science Team Lead, UC Berkeley)*

*Fabienne Bastien (Penn State University)*

*Sarbani Basu (Yale)*

*Thomas Beatty (Penn State)*

*Andrew Bechter (Notre Dame)*

*Eric Bechter (Notre Dame)*

*Cullen Blake (U Penn)*

*Vincent Bourrier (University of Geneva)*

*Bryson Cale (George Mason University)*

*David Ciardi (NASA Exoplanet Science Institute)*

*Jonathan Crass (Notre Dame)*

*Justin Crepp (Notre Dame)*

*Katherine de Kleer (Caltech/MIT)*

*Scott Diddams (NIST)*

*Jason Eastman (Harvard)*

*Debra Fischer (Yale)*

*Jonathan Gagné (U Montreal)*

*Scott Gaudi (Ohio State)*

*Catherine Grier (Penn State)*

*Richard Hall (University of Cambridge)*

*Sam Halverson (MIT)*

*Bahaa Hamze (George Mason University)*

*Enrique Herrero Casas (CSIC-IEEC)*

*Andrew Howard (Caltech)*

*Eliza Kempton (U Maryland)*

*Natasha Latouf (George Mason University)*

*Stephanie Leifer (NASA Jet Propulsion Lab)*

*Paul Lightsey (Ball Aerospace)*

*Casey Lisse (JHU/APL)*

*Emily Martin (UCLA)*

*William Matzko (George Mason University)*

*Dimitri Mawet (Caltech)*

*Andrew Mayo (Technical University of Denmark)*

*Simon Murphy (University of Sydney)*

*Patrick Newman (George Mason University)*

*Scott Papp (NIST)*

*Benjamin Pope (New York University)*

*Bill Purcell (Ball Aerospace)*

*Sam Quinn (Harvard University)*

*Ignasi Ribas (CSIC-IEEC)*

*Albert Rosich (CSIC-IEEE)*

*Sophia Sanchez-Maes (Yale University)*

*Angelle Tanner (Mississippi State)*

*Samantha Thompson (Univ. of Cambridge)*

*Kerry Vahala (Caltech)*

*Ji Wang (Ohio State)*

*Peter Williams (Northern Virginia Community College)*

*Alex Wise (University of Delaware)*

*Jason Wright (Penn State)*

## Table of Contents

## Table of Tables

## Table of Figures

# 1   SCIENCE

## 1.1   EXECUTIVE SUMMARY

> *EarthFinder is a NASA Astrophysics Probe mission concept selected for study as input to the 2020 Astrophysics National Academies Decadal Survey.*

The EarthFinder concept is based on a dramatic shift in our understanding of how PRV measurements should be made. We propose a new paradigm which brings the high precision, high cadence domain of transit photometry as demonstrated by Kepler and TESS to the challenges of PRV measurements at the cm/s level. This new paradigm takes advantage of: 1) broad wavelength coverage from the UV to NIR which is only possible from space to minimize the effects of stellar activity; 2) extremely compact, highly stable, highly efficient spectrometers (R>150,000) which require the diffraction-limited imaging possible only from space over a broad wavelength range; 3) the revolution in laser-based wavelength standards to ensure cm/s precision over many years; 4) a high cadence observing program which minimizes sampling-induced period aliases; 5) exploiting the absolute flux stability from space for continuum normalization for unprecedented line-by-line analysis not possible from the ground; and 6) focusing on the bright stars which will be the targets of future imaging missions so that EarthFinder can use a ~1.5 m telescope.

In this summary we summarize the key findings and recommendations of the report with more detail presented in subsequent sections.

### 1.1.1   *STUDY FINDINGS*

> *"Measurements from space might be a final option if the telluric contamination problem cannot be solved." - National Academies Exoplanet Science Strategy report, 2018*

1. The Earth's atmosphere will limit precise radial velocity (PRV) measurements to ~10 cm/s at wavelengths longer than ~700 nm and greater than 30 cm/s at >900 nm (see Section 1.3), making it challenging to mitigate the effects of stellar activity without a measurement of the color dependence due to stellar activity in the PRV time series. EarthFinder can greatly reduce the effects of stellar jitter through its great spectral grasp, from the UV to the near-IR.

2. Simultaneous visible minus near-infrared (NIR) PRV measurements ("PRV color") perfectly subtracts off the planet signal(s), uniquely isolating the chromatic stellar activity signal from the planet signal(s) in the PRV time-series (Section 1.4.3). EarthFinder's broad spectral grasp offers the highest SNR measurement of this chromatic activity because the lack of the Earth's atmosphere permits PRV measurements at sufficient precision at wavelengths greater than ~700 nm. This unique space advantage will permit disentangling exoplanet and stellar activity signals.

3. "Line-by-line" analysis with high SNR and high-resolution data (R>100,000) can mitigate stellar jitter. In a few cases from the ground, this technique has resulted in a reduction in stellar activity PRV RMS of 33-50% (Dumusque 2018, Lanza et al. 2018, Wise et al. 2018) but greater mitigation (>75%) is needed to detect Earth-mass analogs (Hall et al. 2018). Cegla et al. (2019) demonstrate that with better continuum normalization enabled by a space platform, the ability to distinguish between PRVs and stellar activity from convection and granulation strengthens dramatically (Section 1.4.5).

4.  The UV channel of our space platform permits the simultaneous observations in the near-UV of the Magnesium II lines at 280 nm in addition to the Calcium II H&K absorption lines, the latter of which routinely observed from the ground for PRV activity correlation analysis. These Mg II and Ca II activity sensitive spectroscopic features are produced at different scale heights in the chromosphere of main-sequence Sun-like stars.

5.  Diurnal and seasonal limitations of the ground introduce aliasing which draws power away from the planet signal frequencies and puts them into frequencies that are aliases of one day and one year. EarthFinder provides a large field of regard (FOR) and, for stars outside the FOR, two long visibility windows per year which completely eliminates the diurnal alias and greatly reduces the annual alias (Section 1.4.2). Multiple longitudinally-spaced ground-based telescopes and PRV spectrometers will only partially mitigate daily aliases due to airmass optimization, weather losses and time-varying zero-point velocity offsets between them.

6.  EarthFinder's near continuous observing capability and the efficiency of its diffraction-limited spectrographs give EarthFinder's 1.45 m telescope an effective light gathering power of a much larger ground-based facility.

7.  EarthFinder is perfectly suited to find and characterize the masses and orbits of the planets orbiting ~50 bright main sequence stars (3<V<10 mag) which will be the targets for future NASA flagship missions to image and obtain spectra of nearby Earth-analogs.

8.  High resolving power spectrographs (R~150,000) with simultaneous UV,

visible and NIR coverage offers exciting new capabilities for general astrophysics (Section 1.5).

9.  A preliminary TRL and cost estimate for EarthFinder establishes this mission concept as a Probe-class ($1B) mission with a Kepler-sized telescope using a Kepler-derived spacecraft.

## 1.1.2  STUDY RECOMMENDATIONS

We describe a roadmap for future science and technology work to enable and further refine this mission concept over the next decade:

> *"NASA and NSF should establish a strategic initiative in extremely precise radial velocities (EPRVs) to develop methods and facilities for measuring the masses of temperate terrestrial planets orbiting Sun-like stars."* - National Academies Exoplanet Science Strategy (ESS) report top-level recommendation, 2018

1.  Aligned with the top-level ESS recommendation, we recommend the immediate development of a testbed (e.g. upgrade-able) diffraction limited spectrograph facility with a target single measurement precision and long-term stability of 3 cm/s velocities to investigate the mitigation of stellar and/or solar activity and instrumentation development, to be directly followed by a space PRV mission. It is time now to commence the development of the next generation of PRV spectrometers, testing them on the ground first but also with an application for space. We envision a testbed analogous to NASA JPL's high-contrast imaging testbed facility which combines detailed analysis of error budgets with steady improvements in performance. The facility would require the necessary personnel and science, engineering and technical staff to support its development. This testbed would initially support disk-integrated Solar observations akin to the HARPS Solar telescope feed, so as to correlate and refine the analysis of the high-resolution spectroscopic data with the wealth of information available from heliophysics space and ground assets. This work will be placed into context of the vast wealth of information currently being obtained from visible wavelength seeing-limited spectrometers that are now operating with instrument stability of 10-30 cm/s (e.g. ESPRESSO, EXPRES, NEID), and lay the groundwork for the follow-on space mission EarthFinder. This demonstration must include addressing questions of thermo-mechanical stability under realistic operating conditions for a spacecraft operating at L2. Design and experimental work must be carried out so that each entry in a detailed PRV error budget can be determined with sufficient accuracy so that the overall PRV precision can be predicted.

2.    NASA should convene a workshop to be held by PRV instrument designers, Laser Frequency Comb (LFC) experts, and space electronics engineers to lay out a roadmap for future innovation and TRL maturation. NASA should invest in the development program recommended by these experts. Wavelength standards such as laser frequency combs can reduce the requirement on absolute instrument stability by turning many sources of instrument instability into a common-mode error which can be reduced by reference to a dense, ultra-stable comb of spectral lines. As discussed in Section 2.3, there remains significant work to develop space qualified frequency standards such as laser frequency combs or etalons capable of providing 1 cm/s long term stability over 3-5 years. These frequency standards must provide a dense comb of lines in the visible (0.4-1.0 m) and NIR (1.0-2.5 m) with few GHz spacing so as to be resolvable with EarthFinder's spectrometers.

3.    NASA should invest in a national data analysis center or coordinated funding activity to address the signal processing required to model and mitigate the effects of stellar activity. This effort should comprehensively span the variety of current and future approaches being explored to mitigate stellar activity, including line-by-line analysis, RV color, time-dependent and physically motivated modeling, extreme spectral resolution, etc. to build comprehensive and specialized processing tools and statistical analyses. The scale of the effort required most likely necessitates the specialization of different teams, as opposed to individual PI-led teams attempting to cover all aspects of stellar activity mitigation.

4.    NASA should bridge the NASA Astrophysics division with the extensive expertise in Doppler spectroscopy of the Sun from NASA Heliophysics. In addition to theory and modeling efforts, this includes experiments to extend single-wavelength Solar Doppler observations to space-based and/or balloon-based, multi-wavelength spanning Doppler measurements, and in the NIR free of telluric contamination, with the goal of both understanding our Sun and building better models of stellar activity for mitigating the PRVs of nearby stars with EarthFinder.

**Table 1-1**: The PRV method remains the best technique for measuring planet masses, and the second best for planet discovery after the transit method (only a small fraction of planets, 1-10%, transit their host star).

| Technique / Quantity | Current Performance | Earth-analog Signal Amplitude |
|---|---|---|
| Transit / Radius | ~10 ppm in flux (Kepler, TESS) | ~100 ppm |
| **PRV / Mass** | **~1 m/s (HARPS)** | **9 cm/s** |
| Astrometry / Mass | ~5-16 µas (GAIA, V<12 mag) | 0.03-0.3 µas (@10-100 pc) |
| Direct Imaging / Radius | $10^{-6}$ flux contrast | $10^{-10}$ flux contrast |

## 1.2    STATE OF THE FIELD

The astronomical community is on the cusp of fulfilling the NASA strategic goal to "search for planetary bodies and Earth-like planets in orbit around other stars." (U.S. National Space Policy, June 28, 2010). The 2018 ESS report recommends that "NASA should lead a large strategic direct imaging mission capable of measuring the reflected-light spectra of temperate terrestrial planets orbiting Sun-like stars." (Charbonneau et al. 2018)

> *"The radial velocity method will continue to provide essential mass, orbit, and census information to support both transiting and directly imaged exoplanet science for the foreseeable future." - ESS report top-level finding, 2018*

Without precise radial velocity data, some of NASA's largest planned observatories will fall short of the ultimate goal to determine whether exoplanets can support life. PRVs will provide several critical contributions to the scientific yield and optimization of a future direct imaging mission such as HabEx or LUVOIR which will survey the nearest fifty to several hundred FGKM stars. First, the masses of these planets as determined from PRVs (prior or contemporaneous or otherwise) will be needed for constraining the atmospheric models. Second, the orbits of these planets as determined by PRVs will be necessary to assess habitability. Third, the target selection optimization, observation timing, and required number of direct imaging revisits depend on whether or not we will know a priori from PRVs which nearby stars host Earth-mass planets in Habitable Zone (HZ) orbits.

Currently, the PRV method achieves ~1 m/s single measurement precision with ground-based telescopes searching nearby stars that are known to be quiescent (magnetically inactive) to minimize the negative impact of stellar activity and thus to maximize planet mass sensitivity (see Section 1.4). A new generation of visible wavelength PRV instruments (ESPRESSO, EXPRES, NEID) are on sky now. The best stability demonstrated to date is ~30 cm/s on a time-scale of hours within a single night on a single quiescent target (ESPRESSO SPIE conference presentation, 2018), despite an instrument precision requirement of 20 cm/s. In order to push the sensitivity of the PRV method to 1-10 cm/s on all nearby stars, a space-based PRV would provide a unique platform to overcome many of the factors that challenge the current PRV performance from the ground.

### 1.2.1    SCOPE OF STUDY

> *"Radial velocity measurements are currently limited by variations in the stellar photosphere, instrumental stability and calibration, and spectral contamination from telluric lines. Progress will require new instruments installed on large telescopes, substantial allocations of observing time, advanced statistical methods for data analysis informed by theoretical modeling, and collaboration between observers, instrument builders, stellar astrophysicists, heliophysicists, and statisticians." - ESS top-level finding, 2018*

This final report is the outcome of a partial selection of our proposal submitted in response to the NASA solicitation for the Astrophysics Probes (APROBES) element of NASA's ROSES 2016 (Appendix D.12; NNH16ZDA001N). We were selected to "establish the science case for going to space with a precise radial velocity mission with no funding provided to develop a notional mission architecture or provide mission design lab sessions." (NASA Headquarters selection letter). Consequently, we evaluate the scientific rationale for obtaining PRV measurements in space, which is a two-part inquiry:

- **What can be gained from going to space?** This is addressed in Section 1.4: *Evaluate the unique advantages that a space-based platform provides to enable the identification and mitigation of stellar activity for multi-planet signal recovery in PRV time series.*

- **What can't be done from the ground?** Section 1.3: *Identify the PRV limit, if any, introduced from micro- and macro-telluric absorption in the Earth's atmosphere.*

Many unique additional science cases would also be possible with EarthFinder. We highlight some of these science cases in Section 1.5, including direct exoplanet spectroscopy for characterization, stellar dynamos and asteroseismology, fundamental atomic transitions in the Sun and other stars, following the water in the local Universe obscured by telluric water, and brown dwarf atmospheres.

To assess the technical and programmatic feasibility of EarthFinder, we conducted a short JPL TeamX study to develop an illustrative mission concept to confirm the feasibility and scope of the mission class (Section 4). The TeamX study considered an earlier iteration of the mission concept with a 1.1m primary, but the science case presented herein is for a 1.45m primary (the same as the NASA Kepler probe-class mission). While we were not funded to do an integrated trade between the science yield and mission architecture, the TeamX study establishes the cost for EarthFinder at the top end of a Probe-class mission.

## 1.2.2 *EARTHFINDER OVERVIEW*

> *The primary science goals of EarthFinder are the precise radial velocity (PRV) detection, precise mass measurement, and orbit characterization of Earth-mass planets in Habitable Zone orbits around the nearest FGKM stars.*

These goals correspond to a PRV semi-amplitude accuracy of 1 cm/s on time-scales of several years for a ~10% mass uncertainty (**Figure 1-1**), which can be achieved with 5 cm/s individual measurement precision and taking advantage of binning down the uncertainties from hundreds of measurements. The nominal spacecraft design is discussed in more detail in Section 4, but herein we provide a brief summary.

EarthFinder is based upon the heritage of Kepler spacecraft by Ball Aerospace, with a 1.45-m primary (diffraction limited to ~400 nm). The diffraction-limited beams of starlight are coupled into single-mode fibers illuminating three high-resolution, compact and diffraction-limited spectrometer "arms", one covering the near-UV (280-380 nm), visible (VIS; 380-950 nm) and near-infrared (NIR; 950-2500 nm) respectively with a spectral resolution of greater than 150,000 in the visible and near-infrared arms. The observatory is optimized for the bright (V~5-6 mag) nearby main sequence stars. A small Solar telescope near the solar panels would also be included to obtain Solar spectra.

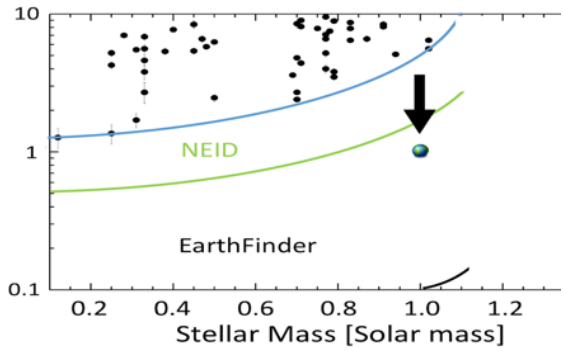

**Figure 1-1**: PRV-discovered exoplanets less than 10 $M_{Earth}$ as a function of stellar mass and planet mass modulo the unknown inclination. Black circles are data from the NASA Exoplanet Archive. The blue-green orb corresponds to the Earth. The blue curve corresponds to the approximate current detection limit of the PRV method, the green curve corresponds to the NEID spectrometer (or similarly, EXPRES, or ESPRESSO), and the black curve corresponds to EarthFinder and its unprecedented 1 cms$^{-1}$ sensitivity.

EarthFinder will be launched into an Earth-trailing (similar to Kepler and Spitzer) or Earth-Sun Lagrange-point orbit. It will have an instantaneous field of regard (FOR) of 70.7% of the celestial sphere, with a continuous viewing zone covering 29% of the sky greater than 45° out of the Ecliptic plane, with 3-6 months of visibility twice per year (e.g. 6-12 months total per year across two "seasons" or visibility windows) for targets within 45° of the Ecliptic plane.

### 1.2.3 *MISSION YIELD*

**EarthFinder will be able to detect Earth-mass planets and their planetary companions around the nearest Sun-like stars.** We carry out a detailed simulation of the yield of a five-year prime mission survey focused on 61 nearby bright stars which would be the likely targets of a future flagship mission to directly image and obtain spectra of Earth-analogs. Stars later than a spectral type of F2 are listed in **Table 1-2**.

We start with the HabEx mission study target list (Gaudi et al 2018), and optimize the target list and exposure times given the

EarthFinder spectrograph spectral grasp, and known stellar properties including coordinates, effective temperature, surface gravity, rotational velocity, apparent magnitude, and metallicity. We use the prescription in Beatty & Gaudi (2015) to estimate the RV precision and the exposure plus readout time necessary to reach a photon detector noise of 3 cm/s across the entire visible and NIR arms. The noise in each arm is >3 cm/s but the velocity uncertainties from both arms add in quadrature to 3 cm/s precision, with the precision higher in the visible arm given the relative RV information content, but the NIR arm providing critical constraints on the stellar activity.

The median target dwell time to reach this precision is 79 minutes. We then add in quadrature a 3 cm/s instrumental error to both arms. We take into account the visibility windows of the spacecraft given an Earth-trailing orbit (similar to Kepler) and known target locations. Each successive target is chosen randomly from the targets visible at a given time to generate a uniform random cadence for minimizing cadence aliases. We include a target slew overhead and assume a dedicated (100% time) survey. **The absolute, long-term stability of the wavelength calibration is <1 cm/s by use of self-referenced laser frequency combs (Section 2.3).**

**Table 1-2**: HabEx Target Catalog used for EarthFinder survey simulations

| Common Name | HIP Number | Spec Type | V mag | Included in Ground Survey? |
|---|---|---|---|---|
| GJ 15 A | 1475 | M2V | 8.13 | Yes |
| GJ 15 B | 1475B | M3.5V | 11.04 | Yes |
| Zet Tuc | 1599 | F9.5V | 4.23 | No |
| Bet Hyi | 2021 | G0V | 2.79 | No |
| 54 Psc | 3093 | K0.5V | 5.88 | Yes |
| HD 4628 | 3765 | K2.5V | 5.74 | Yes |
| eta Cas | 3821 | F9V | 3.44 | Yes |
| 107 Psc | 7981 | K1V | 5.24 | Yes |
| tau Cet | 8102 | G8V | 3.50 | Yes |

**Table 1-2**: HabEx Target Catalog used for EarthFinder survey simulations

| Common Name | HIP Number | Spec Type | V mag | Included in Ground Survey? |
|---|---|---|---|---|
| HD 10780 | 8362 | K0V | 5.63 | Yes |
| HD 16160 | 12114 | K3V | 5.83 | Yes |
| HD 17925 | 13402 | K1V | 6.05 | Yes |
| iot Per | 14632 | G0V | 4.05 | Yes |
| zet01 Ret | 15330 | G2.5V | 5.54 | No |
| zet02 Ret | 15371 | G1V | 5.24 | No |
| kap01 Cet | 15457 | G5V | 4.85 | Yes |
| e Eri | 15510 | G6V | 4.27 | No |
| eps Eri | 16537 | K2V | 3.73 | Yes |
| del Eri | 17378 | K0IV | 3.54 | Yes |
| omi02 Eri | 19849 | K0V | 4.43 | Yes |
| HD 32147 | 23311 | K3V | 6.21 | Yes |
| GJ 191 | 24186 | M1V | 8.853 | No |
| lam Aur | 24813 | G1.5IV | 4.71 | Yes |
| HD 36395 | 25878 | M1.5V | 7.968 | Yes |
| alf Men | 29271 | G7V | 5.09 | No |
| HD 42581 | 29295 | M1V | 8.125 | Yes |
| HD 50281 | 32984 | K3.5V | 6.57 | Yes |
| alf CMi | 37279 | F5IV | 0.37 | Yes |
| 11 LMi | 47080 | G8V | 5.34 | Yes |
| HD 88230 | 49908 | K6V | 6.61 | Yes |
| 36 UMa | 51459 | F8V | 4.82 | Yes |
| HD 95735 | 54035 | M2V | 7.52 | Yes |
| 61 UMa | 56997 | G8V | 5.34 | Yes |
| HD 102365 | 57443 | G2V | 4.88 | No |
| bet Vir | 57757 | F9V | 3.60 | Yes |
| bet CVn | 61317 | G0V | 4.25 | Yes |
| bet Com | 64394 | F9.5V | 4.25 | Yes |
| 61 Vir | 64924 | G6.5V | 4.74 | Yes |
| eta Boo | 67927 | G0IV | 2.68 | Yes |
| HD 122064 | 68184 | K3V | 6.52 | Yes |
| V645 Cen | 70890 | M5.5V | 11.13 | No |
| DE Boo | 72848 | K0.5V | 6.01 | Yes |
| HD 131977 | 73184 | K4V | 5.72 | Yes |
| lam Ser | 77257 | G0V | 4.42 | Yes |
| zet TrA | 80686 | F9V | 4.91 | No |
| 12 Oph | 81300 | K1V | 5.77 | Yes |
| V2215 Oph | 84478 | K5V | 6.34 | Yes |
| 41 Ara | 84720 | G8V | 5.48 | No |
| HD 166620 | 88972 | K2V | 6.40 | Yes |
| sig Dra | 96100 | K0V | 4.68 | Yes |
| bet Aql | 98036 | G8IV | 3.71 | Yes |
| del Pav | 99240 | G8IV | 3.56 | No |
| HD 191408 | 99461 | K2.5V | 5.32 | No |

**Table 1-2**: HabEx Target Catalog used for EarthFinder survey simulations

| Common Name | HIP Number | Spec Type | V mag | Included in Ground Survey? |
|---|---|---|---|---|
| HD 192310 | 99825 | K2V | 5.723 | Yes |
| 61 Cyg A | 104214 | K5V | 5.21 | Yes |
| 61 Cyg B | 104217 | K7V | 6.03 | Yes |
| AX Mic | 105090 | M1V | 6.68 | No |
| eps Ind | 108870 | K5V | 4.69 | No |
| TW PsA | 113283 | K4V | 6.48 | No |
| HD 217987 | 114046 | M2V | 7.34 | No |
| HD 219134 | 114622 | K3V | 5.57 | Yes |
| **Targets not included in space mission simulation:** | | | | |
| ups And | 7513 | F9V | 4.10 | Yes |
| del Tri | 10644 | G0.5V | 4.87 | Yes |
| tet Per | 12777 | F8V | 4.11 | Yes |
| pi.03 Ori | 22449 | F6V | 3.19 | Yes |
| gam Lep | 27072 | F6V | 3.60 | Yes |
| HD 103095 | 57939 | K1V | 6.45 | Yes |
| gam Ser | 78072 | F6V | 3.84 | Yes |

We compare our EarthFinder simulated survey to a ground-based survey yield for a subset of 44 of the 61 space targets, plus 7 additional targets (51 total) accessible from a Northern Hemisphere facility. We assume a "super"-NEID, also capable of achieving 3 cm/s instrument stability, but with the wavelength coverage, efficiency and resolution of NEID. We place the super-NEID on the LBT, an 8-m class telescope. We account for the properties of the spectrograph including spectral resolution, spectral grasp and throughput efficiency, detector noise and read out times, as well as the slew-rate and pointing limits of the telescope. We account for target airmass and hour angle to optimize sequential target selection to minimize airmass and to minimize the time since a given target was last observed, based upon the MINERVA survey dispatch scheduler (Newman et al. in prep.). We also account for the realistic weather losses, sunrise and sunset times for Arizona. We assume a PRV survey using 25% of the total nights available for five years.

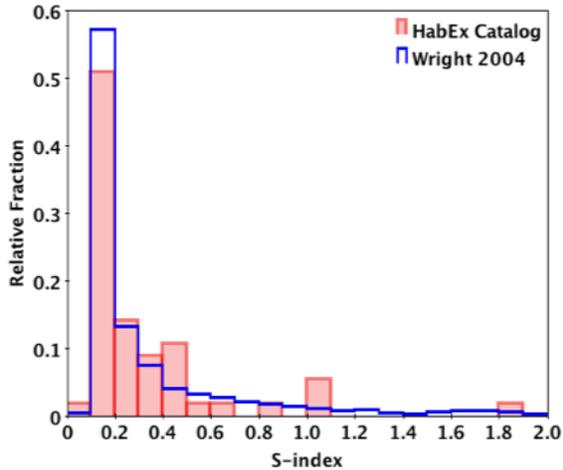

**Figure 1-2**: HabEx targets are on average more active than the typical quiescent PRV survey target. We compare a histogram of S-index activity indicators for stars in the HabEx catalog (red) compared with the S-index distribution of the California Planet Search survey sample (blue; from Wright et al. 2004).

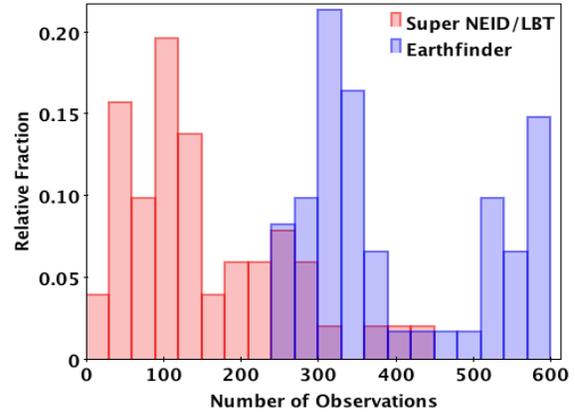

**Figure 1-3**: A dedicated EarthFinder mission survey of 61 direct imaging targets with a 1.45m aperture (blue) **outperforms** a 25%-time survey on a 8-m ground-based facility of 53 targets (red) in the number of observations per star, with a median of 349 vs. 124 epochs respectively.

correction of stellar activity and tellurics in Sections 1.4 and 1.3 respectively.

We inject a simulated Universe of planets into our simulated RV survey data using the SAG13 demographics[1]. We follow a random draw prescription for the mass and period of the exoplanets that is nearly identical to the exoplanet demographics used for the HabEx mission yield calculations based upon the nominal SAG13 occurrence rates (Dulz et al. in prep). We remove less massive planets without replacement that are within a mutual separation of 9 Hill radii of a more massive planet (Lissauer et al. 2011). This results in a range of 1-9 planets per star, with an average of 5-6 planets per star. We use the FORECASTER mass-radius relation of Chen & Kipping (2017). We randomly draw system inclinations and longitudes of periastron, and we randomly draw eccentricities from a Beta distribution following Kipping (2013).

For this initial comparison of EarthFinder and ground-based surveys, we assume that stellar activity and the effects of telluric atmospheric lines are perfectly corrected in our simulated surveys (both ground and space). We address the impact of the imperfect

We examine the ability to recover orbital parameters of each planet in the system in the two datasets. Using a custom-modified version of the RADVEL analysis software that makes use of the recent emcee v3.0rc2 MCMC-sampler Python library (Fulton et al. 2018), we evaluate the Bayesian posterior probability distribution evidence supporting the existence of each planet in the PRV time-series data for a given system, performing a model comparison and evaluating the log-likelihoods with all combinations of planets removed. Statistically favored models are considered recovered if the periods and velocity semi-amplitudes match the injected parameters within a factor of <50%, although often the match is much better, and are considered false positives otherwise. The other planets in the system are either noted as marginally recovered or excluded detections.

We compare the EarthFinder orbital parameter recovery to our ground-based survey. EarthFinder outperforms the ground-based survey by factors of 3 on average in the precision of the recovered velocity semi-amplitude for

[1] https://exoplanets.nasa.gov/exep/exopag/sag/#sag13

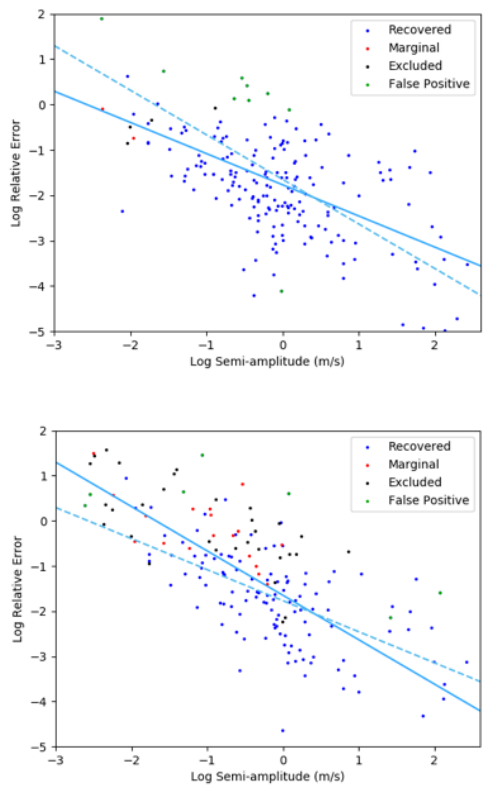

**Figure 1-4: The cadence advantage of EarthFinder enables it to detect Earth-mass HZ analogs with three times better accuracy on average for the recovery of the velocity semi-amplitude relative to an equivalent ground-based survey. Once factoring in the telluric and stellar activity correction error terms present from the ground, the space advantage grows.** Top: Log of the absolute relative error in the recovered vs. injected velocity semi-amplitude as a function of the semi-amplitude for the simulated planets in the EarthFinder survey. A linear fit is shown as a solid line. Bottom: The same for the ground-based survey. The dashed lines are the solid lines from the other panel, overlaid for a direct comparison to highlight the space advantage in performance.

Earth-mass planets. This is due to the advantages of EarthFinder's cadence, which greatly reduces aliasing (see **Figure 1-15** in Section 1.4.2).

> *At a velocity semi-amplitude for a Earth-mass planet in a HZ orbit (9 cm/s), the cadence advantage of EarthFinder provides a factor of three on average improvement in accuracy compared to the ground-based survey (average relative fractional error in the velocity semi-amplitude of ~10% for EarthFinder vs ~30% from the ground).*

## 1.3 EARTHFINDER ELIMINATES TELLURIC CONTAMINATION

The assumption of perfect atmospheric correction and quiescent stars made in the previous section is, of course, optimistic and incorrect. Spectral contamination due to the absorption lines of Earth's atmosphere (telluric absorption) poses a serious challenge to PRVs. Unlike atmospheric emission line contamination, these absorption features cannot be removed through sky subtraction during spectral image reduction. It is a known bottleneck for achieving higher RV precision in the NIR (Bean et al. 2010). Moreover, it was recently realized that even the "micro-telluric" lines (flux depths <2% and mostly <1%) at visible wavelengths can contribute to RV error budget at 20-50 cms$^{-1}$ (Cunha et al. 2014; Artigau et al. 2014; Wang et al. 2019). This is a large term in the PRV error budget which is eliminated in space.

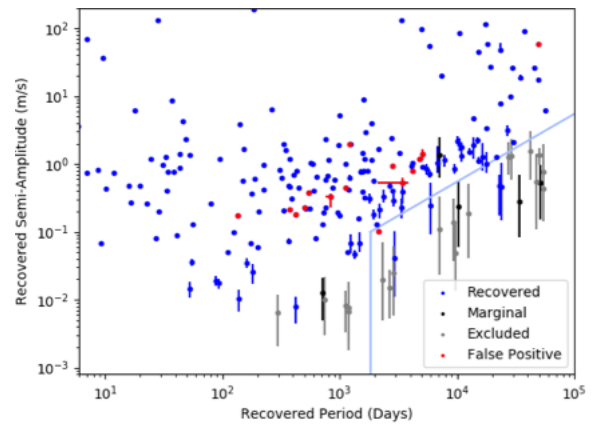

**Figure 1-5**: **EarthFinder detects most (>90%) of our simulated exoplanets, including most of the 23 Earth-mass HZ analogs with semi-major axes 0.95<a<1.67 and masses 0.5<m<4 m$_{Earth}$** (following Kopparapu et al. 2013). We do not test the recovery of all planets with orbital periods >5 years and K(m/s)/P(day) < 5.476*10^-3 (m/s/day) shown as the blue line for computational efficiency. Detecting more Earth-mass planets requires an intrinsically higher value of η$_{Earth}$ or surveying more stars with a larger aperture or longer primary mission. By comparison, GAIA will only be able to detect the Jovian planets greater than 1 Jupiter mass in the upper right corner of this plot (K >10 m/s, approximately).

Even at low airmass, the impacts of micro-telluric lines are ~20 cms$^{-1}$. These shallow but prevalent lines are challenging to model due to time-variability and a lack of accurate laboratory measurements of water lines. It is currently unclear how much we can eliminate their impacts on PRVs beyond the 0.5 m/s precision level (Fischer et al. 2016).

The situation is much worse in the NIR, where deep and saturated telluric lines leave only several small clean spectral windows. The saturated lines completely block our access to the spectral windows near 1.4 and 1.9 μm, and the deep lines are very difficult to model telluric lines to a precision of 1-2% or better (e.g., Seifahrt et al. 2010, Gullikson et al. 2014, Smette et al. 2015). A 1-2% modeling residual would contribute to the RV error budget at ~0.5 m/s level, similar to the micro-telluric lines in the optical. For example, a study by Sithajan et al. (2016) characterized the effects of telluric contamination and effectiveness of some typical remedies (masking and modeling) for emission line-calibrated spectra for the optical, broad optical (300-900 nm), and NIR. They concluded that, even if all of the telluric lines are modeled and subtracted to the 1% level, the residuals would still cause 0.4-1.5 m/s RV errors in the NIR for M and K dwarfs.

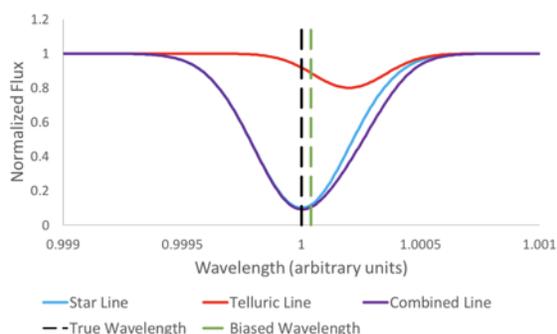

**Figure 1-6**: Convolved telluric lines "pull" the centroid of stellar absorption lines off from their true Doppler-shifted stellar absorption lines, as shown in this illustration (Wright et al. 2013).

Modeling the tellurics to 1-2% residuals is challenging; state-of-the-art works typically achieve 2-7%. Sameshima et al. (2018) achieves residuals of 2% for water lines in 0.9-1.35 μm using telluric standard star observations. However, they note that the time and spatial variability of water absorption puts stringent requirements on the telluric standard star observations. Ulmer-Moll et al. (2019) compared several methods and also most commonly used software packages to correct for telluric absorption lines using synthetic spectra, and their best results typically have residuals around 3-7% (1-2% at the very best).

The challenges posed by telluric contamination are recognized by the community, as summarized in the ExoPAG white paper by Plavchan et al. (2015) and the review of the field presented as a result of the second Extreme PRV workshop (Fischer et al. 2016).

> *Telluric contamination is one of the largest error budget terms in ground-based PRV surveys (Halverson et al. 2016), and it can only be completely eliminated by getting above the Earth's atmosphere. EarthFinder will have unimpeded access to the entire visible and NIR spectral range in a single shot.*

We perform simulations with synthetic spectra and extract RVs from these spectra to assess the RV precision limit set by the telluric contamination for ground-based instruments. With our controlled simulations, we isolate the effects of tellurics from other factors such as photon noise, stellar activity/jitter, and instrumental effects. Our goals are: (1) to quantify the RV precision limit set by the telluric contamination in a broad wavelength range, from 350 nm to 2.5 μm; (2) to characterize the effectiveness of commonly used ("Division" & "Modeling") methods for mitigating tellurics.

### 1.3.1 METHODOLOGY

We simulate observed spectra at an R=120,000 over the course of one year (365 nights) on the Sun using the Kurucz solar spectrum generated by ATLAS9 (Kurucz et al. 2005). We add telluric absorption in each simulated spectrum using synthetic telluric spectra generated by TAPAS (Bertaux et al. 2014). Overall, our simulation aims to reproduce an ensemble of realistic spectra of a Sun-like star through the Earth's atmosphere throughout a year, and we then extract RVs from these spectra using different telluric mitigation methods and compare their effectiveness. For each night, we randomly assign an airmass representing a typical target airmass distribution, and we randomly draw a value for the amount of precipitable water vapor representative of PWV values from Kitt Peak (PWV; unit is mm; https://www.suominet.ucar.edu/).

We extract RVs from the simulated spectra with three different methods:

1. <u>No correction:</u> RVs are extracted via computation of a cross-correlation function (CCF) method following Baranne el al. (1996). We assume a perfectly known spectral point spread function (PSF), and wavelength solution, which are not unrealistic for the ultra-stabilized next-generation spectrographs calibrated by laser frequency combs. We also use a perfectly known stellar template to perform the cross-correlation (CCF), which is unrealistic but allows us to isolate the effects of tellurics.

2. <u>Division:</u> RVs are extracted via a CCF but before the CCF is computed, a telluric spectrum is removed via mathematical division. The telluric spectrum being divided out is exactly the same one as used for generating the simulated spectrum for each night and convolved to R=120,000 in the same fashion. This approach is the most commonly used method to mitigate the effects of tellurics in PRVs for stabilized spectrographs such as HARPS and CARMENES.

3. <u>Modeling:</u> RVs are extracted via the forward modeling method similar to Butler et al. (1996). We again assume a perfectly known wavelength solution and spectral PSF. Instead of using the telluric spectra exactly the same as the ones synthesized into the observed spectrum just like in the Division method, we use a different set of telluric spectra with the wrong line profiles as the model input to mimic our lack of knowledge of the atmospheric conditions and molecular line profiles. This set of telluric spectra were generated by TAPAS using atmospheric conditions of Mauna Kea (instead of Kitt Peak). **Figure 1-7** illustrates this mismatch of line profile.

When extracting RVs, we divide each spectrum into 230 chunks, with chunk size growing linearly as the wavelength from 0.3 nm at 350 nm to ~20 nm at 2.5 μm.

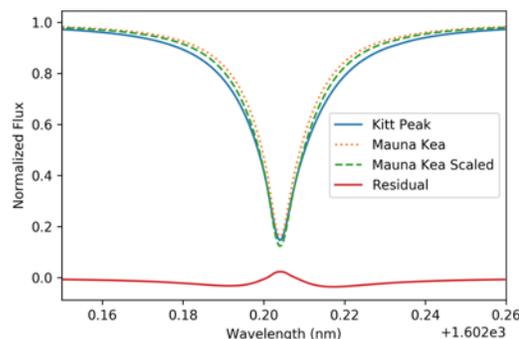

**Figure 1-7**: Comparison of the line profiles for a $CO_2$ line near 1602 nm for two observatories: Kitt Peak (used in simulated spectra) and Mauna Kea (used as input to fit the simulated spectra). Both spectra are at airmass = 1. The Mauna Kea Scaled spectrum is the best-fit version when fitting the Kitt Peak profile by scaling the Mauna Kea line by a power law. The RMS of the residual of this fit is 1.2%, which is around the typical value for all lines and smaller than the typical values reported by Ulmer-Moll et al. (2019).

## 1.3.2 ERRORS INDUCED BY TELLURICS VS. WAVELENGTH

> *When no corrections are applied, tellurics induce considerable amounts of errors from cm/s to more than km/s for different spectral regions.*

**Division or Modeling effectively removes some of the RV errors induced by tellurics, but not completely. Figure 1-8** illustrates the RV errors as a function of wavelength for RVs extracted with the three methods described (smoothed across wavelength for visualization purposes). In the visible, the Division method effectively mitigates the errors added by tellurics, bringing them down by an order of magnitude. These results represent an idealized situation and thus a lower limit for the RV errors, since it hinges on the perfect knowledge of the spectral continuum and such perfect knowledge is unrealistic for ground-based instruments.

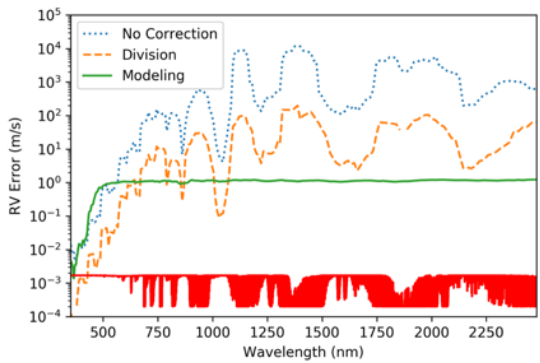

**Figure 1-8**: RV errors added by tellurics as a function of wavelength for three different methods used in this work. Each point plotted is the average RV error for 7 neighboring chunks centered at each wavelength. The RV error of each chunk is the RMS of RVs of this chunk over the simulated time span of 365 days. The spectrum plotted in red at the bottom is an illustration of telluric absorption.

For the NIR with moderate to deep telluric absorption, Modeling outperforms Division significantly. This is because division is not mathematically correct. As the star-light goes through the atmosphere first and is then broadened by the spectrograph, mathematically, this corresponds to a multiplication of the stellar spectrum with the telluric spectrum and then a convolution with the spectral PSF. As convolution is not distributive over multiplication, dividing out the broadened telluric spectrum from the observed spectrum will induce an additional error that is proportional to the telluric line depth. Therefore, the Division method does not work as well in regions with moderate or deep telluric lines.

## 1.3.3 TELLURICS' CONTRIBUTION TO THE RV ERROR BUDGET

> *Through our simulations, the optimistic floor of RV precision from the ground due to the telluric contamination is around 2 cm/s in the visible (<700 nm) and 30 cm/s in the red/NIR.*

To quantify the contribution to the RV error budget by tellurics for different bandpasses or under the contexts of different instruments, we combine the RVs from different spectral chunks to report a final RV for each night, and compute the RMS of these nightly RVs as the final "RV error". Table 3 gives the final RVs for different wavelength ranges that correspond to several state-of-the-art RV instruments, including an EarthFinder equivalent (EFE) instrument on the ground.

The RV errors listed in **Table 1-3** represent optimistic lower limits for the errors induced by the telluric contamination. Our simulations have many idealized assumptions in order to isolate the effects of tellurics in the observed RV spectra, including perfect knowledge of the telluric spectrum on any given night, and a perfect continuum correction. Under realistic conditions, changes in atmospheric conditions during and between exposures will affect the continuum absorption by tellurics, which would translate into additional RV errors. As a result of these and other assumptions, the floor of 30 cm s$^{-1}$ is an optimistic estimate, and the errors caused by tellurics in the NIR may be larger.

**Table 1-3**: Summary of RV rms precision lower limits for different instrument using different RV extraction methods. EFE stands for EarthFinder Equivalent on the ground.

| Ground-based Instrument | No Correction (m/s) | Division (m/s) | Modeling (m/s) |
|---|---|---|---|
| EFE Visible Arm (380-900 nm) | 0.035 | 0.021 | 0.067 |
| EFE NIR Arm (900-2500 nm) | 2.432 | 0.761 | 0.321 |
| ESPRESSO (380-788 nm) | 0.034 | 0.020 | 0.069 |
| EXPRES (380-680 nm) | 0.031 | 0.013 | 0.068 |
| NEID (380-930 nm) | 0.034 | 0.019 | 0.067 |
| CARMENES Visible Arm (520-960 nm) | 0.169 | 0.098 | 0.194 |
| CARMENES NIR Arm (960-1710 nm) | 2.359 | 0.659 | **0.442** |

## 1.3.4   ADDITIONAL LIMITATIONS SET BY TELLURICS

Several additional PRV error terms are introduced by telluric contamination that are not included in our simulations above.

> *First, telluric contamination induces spurious signals in the RVs at periods that are harmonics of one year, which is particularly vexing when searching for HZ Earth-mass analogs.*

1. The line pulling effects caused by tellurics that induce RV biases and errors are correlated with the barycentric velocity, which is due to Earth's annual motion. **Figure 1-7** and **Figure 1-9** illustrate this effect, which impacts our ability to detect Earth-like planets. This effect is most pronounced for the Division method because the residuals of the division correlate strongly with telluric lines. As a result, RVs from the Division method exhibit strong periodogram peaks at periods that are harmonics of one year.

Second, changes in the spectral continuum because of changes in the telluric absorption and Rayleigh scattering introduce additional RV errors. The changes in the spectral continuum can be hard to model out or account for accurately, in part because any residual slope or curvature in the continuum translates into biases and errors in the RV measurements.

Third, **deep and saturated telluric absorption cause significant loss of RV information content in the NIR**. This affects the RV precision in two ways: (1) For regions with saturated telluric lines, or even just dense and deep telluric lines, it is prohibitive to construct a high-fidelity stellar mask for CCF or a perfect stellar template for forward modeling. (2) The blocked spectral windows by tellurics often line up with the stellar spectral region that are rich in molecular absorption lines, especially in M/K dwarfs (see **Figure 1-10**). This causes a significant loss of RV information content, fundamentally limits the NIR precision, and thus our ability to mitigate stellar activity (**see Section 1.4.3**).

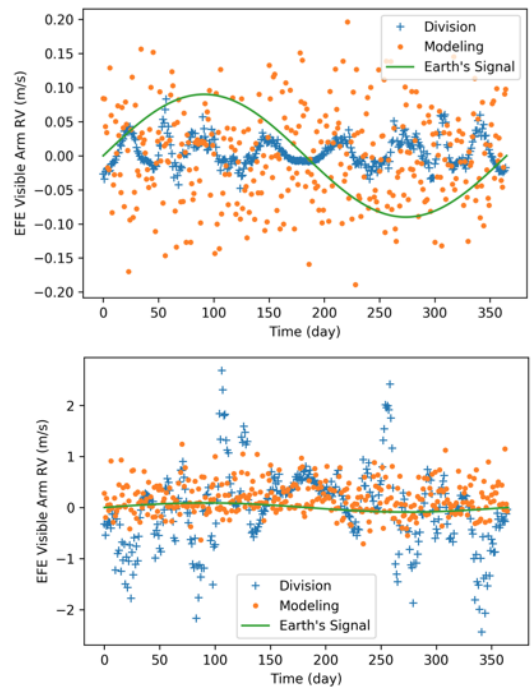

**Figure 1-9**: RV signals vs. time, as measured by the visible arm (upper panel) and NIR arm (lower panel) of an EarthFinder equivalent (EFE) spectrograph from the ground, extracted with two different telluric mitigation methods, assuming SNR=100 per pixel for R=120,000. The RV signal of an Earth analog is plotted as the green line (semi-amplitude = 9 cms$^{-1}$).

## 1.4 EARTHFINDER CAN UNIQUELY MITIGATE STELLAR JITTER

### 1.4.1 *STELLAR SIGNALS*

Below 1 m/s precision, RV measurements are affected by different types of stellar signals at different timescales. Those signals perturb the detection of small-mass planets. The Earth induces a Doppler signal of 9 cm s$^{-1}$ on the Sun, an effect that has an amplitude an order of magnitude smaller than the known stellar signals. The known stellar signals, with their amplitudes and timescales, are summarized in **Table 1-4**.

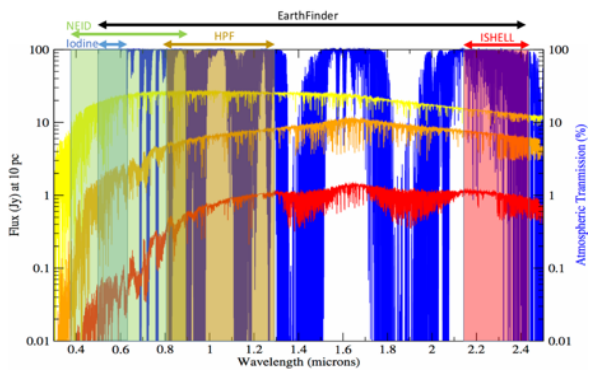

**Figure 1-10:** Telluric absorption spectrum (blue) plotted as a function of wavelength, and compared with G, K and M dwarf spectra in yellow, orange and red respectively, showing the overlap of spectral regions rich in stellar RV information content blocked by deep telluric absorption.

*EarthFinder will have the highest SNR measurements available from the near-UV to the NIR, with extremely high-resolution spectra, and with a near-perfect cadence sampling. Together, this will allow us to characterize stellar signals like never before with a variety of approaches including the RV color, line-by-line analysis, cadence, and simultaneous photometry, to mitigate the impact of stellar activity down to a level of a few dozens of cm/s, therefore enabling the detection of Earth analogs.*

**Stellar oscillations**: Oscillations in solar-type stars are induced by pressure waves that drive a dilatation and contraction of external envelopes on timescales of a few minutes (Schrijver & Zwann 2000, Broomhall et al. 2009). The amplitudes of these oscillation modes range from a few dozens of cm s$^{-1}$ to a few m s$^{-1}$ for solar-type stars (Dumusque et al. 2011a, Bedding & Kjeldsen 2007). Because of the short timescale and narrow frequency range of stellar oscillations, it is possible to average out this perturbation down to a few dozen of cm s$^{-1}$ by increasing the exposure time to at least the period of the signal (Dumusque et al. 2011a, Chaplin et al. 2019).

**Flares**: On the timescale of approximately an hour, stellar flares can influence the RV of very active stars (Reiners et al. 2009). However, for stars like the Sun, Saar (2009) estimated that the effect of flares should be smaller or equal to 0.5 m s$^{-1}$ for moderate flares. In any case, it is possible to probe the core of some sensitive lines to probe flares and discard the affected measurements.

**Table 1-4**: Known sources of stellar signal, that EarthFinder can model and mitigate, with their typical timescales and amplitudes for main sequence stars.

| Stellar signal | Timescale | Amplitude | References |
|---|---|---|---|
| Oscillations | <15 min | ~1 m/s | Kjeldsen et al. 95, Bouchy & Carrier 01, Butler et al. 04, Bedding & Kjeldsen 07 |
| Flares | ~1h (only active M) | <0.5 m/s | Saar 09, Reiners 09 |
| Granulation | 5 min- 2 days | ~1 m/s | Del-Moro et al. 04, Del-Moro 04, Cegla et al. 13, Cegla et al. 14 |
| Short-term activity (spots, faculae) | 10-100 days (stellar rotation) | a few m/s | Saar & Donahue 97, Queloz et al. 01, Meunier et al. 10, Aigrain et al. 12, Dumusque et al. 14, Meunier et al. 17 |
| Grav. redshift | 10 days - 10 years | <0.1 m/s | Cegla et al. 12 |
| Long-term activity (Magnetic Cycles) | ~10 years | 1-20 m/s | Makarov 10, Dumusque et al. 11, Meunier et al. 13 |

**Granulation**: The surfaces of Sun-like stars are composed of a convective envelope, where hot bubbles of gas, known as granules, rise to the surface, and eventually cool and sink back down into intergranular lanes. To an observer, the rising plasma is blueshifted, whilst the sinking plasma is redshifted. Some of the up- and down-flows cancel out, but this still leaves net RV shifts that vary at the level of several 10s of cm s$^{-1}$. From the short-term small convective pattern of granulation (Title et al. 1989) to the long-term very large structure of super-granulation (Roudier et al. 2014, Del Moro 2004), granulation phenomena affect RV measurements on minutes to a few days timescales, with amplitude of ~30-70 cm s$^{-1}$ (Elsworth et al. 1994; Pallé et al. 1999). To average out the signal from granulation, an efficient approach used so far is to observe a star several times during the same night, and nightly bin the data (Dumusque et al. 2011). Meunier et al. (2015) showed however that granulation signals can be averaged out after binning over several days. This may partially be due to the fact that the granules tend to appear and disappear in the same locations; hence, even if we average over periods of time much greater than the lifetimes of the granules we are still unable to completely bin out the granulation impact. This can be seen on the Sun, where we are able to resolve the granulation patterns, which are still clearly visible even after averaging exposures together for an entire hour (>10 times the lifetime of a single granule; Cegla et al. 2019). Work from Cegla et al. (2014) and Cegla et al. (2019), based on magneto-hydrodynamic simulations of the Sun, show, however, that the bisector variation of spectral lines is strongly correlated to the RV effect of granulation and it should be possible to mitigate this perturbing signal down to 0.1 m/s (Section 1.4.5).

**Short-term activity (spots, faculae)**: On longer timescales, e.g. stellar rotational periods, the presence of active regions, spots or faculae on the stellar surface perturb PRV measurements. Active regions induce RV variation by (1) breaking the flux balance between the blue-shifted approaching limb and the red-shifted receding limb of a rotating star (e.g. Lagrange et al. 2010, Desort et al. 2007, Saar & Donahue 1997) and by (2) breaking the RV balance of the convective blueshift, because active regions appear redshifted compared to the quiet photosphere due to the local inhibition of convective blueshift by strong magnetic fields (Dumusque et al 2014, Meunier et al. 2010, Cavallini et al 1985, Dravins et al. 1981, Haywood et al. 2016). Due to its nature, short-term activity will induce spectral line shape variations with a timescale of the stellar rotational period. This will appear in the RVs as quasi-periodic signals at the rotational period of the star and its harmonics, which can be fitted with sine waves to mitigate the RV activity signal (Boisse et al. 2011, Queloz et al. 2009). In some cases, where simultaneous RV and photometric data are available, it is possible to use the information of the photometry alone, to de-correlate the RV measurements from short-term activity signal using the FF' method (Haywood et al. 2014; Aigrain et al. 2012). Probably the best methods used thus far is to model the correlated signal induced by the short-term activity signal using Gaussian Processes (GPs) or a Moving Average (e.g. Haywood et al. 2014, Damasso et al. 2018, Jones et al. 2017, Grunblatt et al. 2015, Rajpaul et al. 2015, Tuomi et al. 2013a, Tuomi et al. 2013b). However, it seems that these optimal methods are not yet successful in recovering planetary signals with amplitudes smaller than 0.5 m/s (Dumusque et al. 2017), and therefore cannot be used to characterize Earth-twins. Another method to mitigate short-term stellar activity is to observe stars in the NIR as the flux contrast of active regions is smaller at longer wavelengths. The signal from short-term stellar activity is thus chromatic and smaller in the NIR.

Therefore, by comparing the visible and NIR RV measurements, it is possible to de-correlate the chromatic activity signal from the grey planetary signal (Zechmeister et al. 2018, Marchwinski et al. 2015, Huélamo et al. 2008; see Section 1.4.3).

**Long-term activity (magnetic cycles)**: Finally, over the long-term, RV measurements are affected by magnetic cycles (e.g. Lovis et al. 2011, Lindegren et al. 2003). Because more and more active regions appear on the stellar surface, the total inhibition of convection produced by the presence of active regions increases, the star appears more redshifted, and therefore a positive RV variation is observed. Magnetic cycles of stars induce a RV effect of a dozen of m/s over several years in the case of the Sun (Lovis et al. 2011). This is an important challenge to the detection of Jupiter analogues, but also for tiny planetary signals at shorter periods, as small errors when fitting long-term trends can create spurious signal at short periods due to aliasing (Rajpaul et al. 2016). The RV effect induced by magnetic cycles is well correlated with activity indicators like the log (R'HK), and a simple de-correlation can significantly reduce this perturbing signal (e.g. Delisle et al. 2018, Diaz et al. 2016).

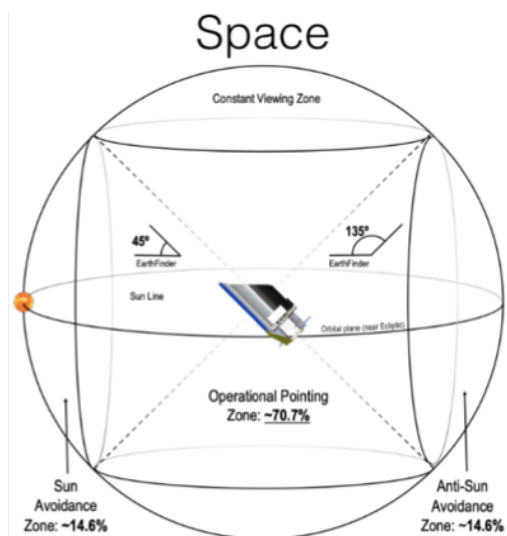

**Figure 1-11**: The wide field of regard of EarthFinder is superior to ground-based facilities which are additionally limited by weather.

> *Among all stellar signals, the most significant and difficult ones to correct for from the ground is short-term activity and granulation. Those signals need to be mitigated down to a level of a few dozens of cm/s if we want to be able to characterize an Earth analog. It is only with a combination of perfect sampling, near-UV to NIR capabilities and extremely high SNR and resolution that EarthFinder provides to make this correction possible. EarthFinder will be the only instrument providing all of these characteristics at once and therefore will have the highest probability of detecting Earth analogs.*

## 1.4.2 THE ADVANTAGE OF HIGH CADENCE IN PLANET DISCOVERY AND MITIGATION OF STELLAR ACTIVITY

EarthFinder's wide Field of Regard (**Figure 1-11**) allows excellent sampling of the RV time series which is essential to avoiding aliases and degeneracies in the search for exoplanets. As discussed in Section 1.4.1, the RV signal measured from a stellar spectrum is the sum of signals from a number of stellar surface activity features modulated both by the rotation period of the star and the long magnetic cycle, with the stellar reflex motion from the planetary orbits added into this. Thus, the RV signal has a number of frequency components which may be periodic or quasi-periodic. Interpreting this signal in an unambiguous manner places a minimum requirement on the sampling of that signal.

> *EarthFinder will have a large continuous viewing zone with an orbit that provides two viewing windows per year, completely eliminating the diurnal alias and eliminating or reducing the annual alias. It will therefore be the best instrument to achieve optimal sampling, which will allow us to differentiate between stellar and planetary signals, particularly at HZ orbital periods of ~0.5-2 years.*

EarthFinder's sampling can be tailored to maximize signal retrieval within the desired discovery space of parameters. From the ground, the number of observations is limited by day/night cycles, weather outages and sky visibility due to the time of year.

To demonstrate the benefit of improved cadence compared to the ground-based status quo we present selected results from the recently published work of Hall et al. (2018). The simulations from this paper are not completely analogous to the EarthFinder case; they were performed for HARPS3, a future ground-based high-resolution spectrograph. The photon noise per data sample is an order of magnitude higher for HARPS3 compared to EarthFinder (30 cm s$^{-1}$ compared to 3 cm s$^{-1}$) and a simple schedule of one data point per 24 hours is taken as the baseline for the uninterrupted (i.e. space-based) dataset. The primary purpose of this was to demonstrate the benefit of no gaps due to weather and yearly sky visibility windows that are the natural limitations of ground-based observations. RVs are simulated over two time periods (5 years and 10 years) for a number of planetary system architectures (on Keplerian orbits) using a stellar mass of 0.8 M$_\odot$ (~G5 dwarf star). White noise is added at the 30 cm s$^{-1}$ level.

A second dataset is also generated that contains some simulated stellar noise signals. The stellar activity related RV signals are generated using Spot Oscillation and Planet (SOAP) 2.0 model (Dumusque et al. 2014) and added to the planetary RV signals with white noise. The limitations of this model are that it does not include the effects due to granulation, pressure-mode oscillations and variations due to the long-term magnetic cycle. The focus of the stellar activity model is on the contribution due to spots and faculae: their signals are modulated on the rotation period of the star and evolve with a typical lifetime. For the RV analysis in Hall et al. (2018) it is assumed that we can suppress the stellar activity signal by a factor of four (75% level correction). This level of correction has yet to be demonstrated. However, as stated in Section 1.4.4, very recent analysis shows that by studying the behavior of individual spectral lines, we are able to mitigate stellar activity down by nearly a factor of two. Furthermore, there is still room for improvement, particularly from space.

A Bayesian analysis method is used for model comparison (i.e. the likelihood of the data being best described by a 0, 1, 2, or 3 planet model) and the best fit parameters for all the planets within each of those models. System-2 from Hall et al. (2018) is a 3-planet system and is analogous to our own Solar System, containing two rocky planets (like Venus and Earth) and a Jupiter-like planet; the planet parameters are listed in **Table 1-5**. The results of the 3-planet solution are shown in **Figure 1-12**. The results using the 5-yr space schedule identifies the Earth twin successfully; it also finds the Venus analogue with good estimates for the period and RV amplitude, with just the phase slightly off. The results using the 10-yr space schedule finds all three planets.

From the ground, multiple longitudinally-spaced telescopes can be used to partially mitigate daily aliasing. However, weather losses, and the time-varying zero-point velocity offsets between the different telescope sites must be modeled and will still introduce errors and daily aliases. Space-based observations largely eliminate these difficulties.

*The results demonstrate that optimized sampling is crucial in the retrieval of Earth-twin RV signals. Such sampling is much more easily achieved from space.*

**Table 1-5**: List of the planetary parameters used in the System-2 model.

| Planet Mass (M$_{Earth}$) | Period (d) | RV (m/s) |
|---|---|---|
| 0.82 | 197 | 0.11 |
| 1.00 | 293 | 0.11 |
| 200 | 2953 | 10.34 |

### 1.4.3 THE RADIAL VELOCITY COLOR ADVANTAGE OF EARTHFINDER

Planet RV signals are achromatic: the same velocity reflex motion is measured at all wavelengths. Conversely, RV variations due to stellar activity are chromatic, particularly the most vexing effects from spots and faculae, since the flux emission from the stellar surface is temperature and thus wavelength dependent (Reiners et al. 2009, Tal-Or et al. 2018, Zechmeister et al. 2018, Wise et al. 2019). The signal in one wavelength regime (e.g. the blue or visible) will be a summation of the planetary signals and the stellar activity, which will be different in a second wavelength regime (e.g. the red or NIR), a signal that is also a summation of the planetary signals and a modified stellar activity signal. Thus, by simultaneously measuring RVs in two different wavelength regimes and then subtracting these two time-series (e.g. the RV color time-series of blue minus red, or visible minus NIR), the planet signals subtract out perfectly, leaving only the chromatic activity signal.

> No other technique for mitigating stellar activity besides simultaneous measurements of RV color allows for the perfect isolation of the activity signal from the planet signals. EarthFinder, by virtue of spanning the largest spectral range, offers the highest SNR determination of RV color superseding any ground-based facility.

The measurement of simultaneous RV color - e.g. measuring the RV in two different wavelength regimes, bands or spectrograph arms at the same time - is relatively noisier when restricted to visible wavelengths due to the limited wavelength difference. **Since the chromaticity of activity goes as $1/\lambda$ to first order, the RV color amplitude goes as the amplitude of the activity in one wavelength regime times the fractional wavelength difference between the two RV measurements in each wavelength regime.**

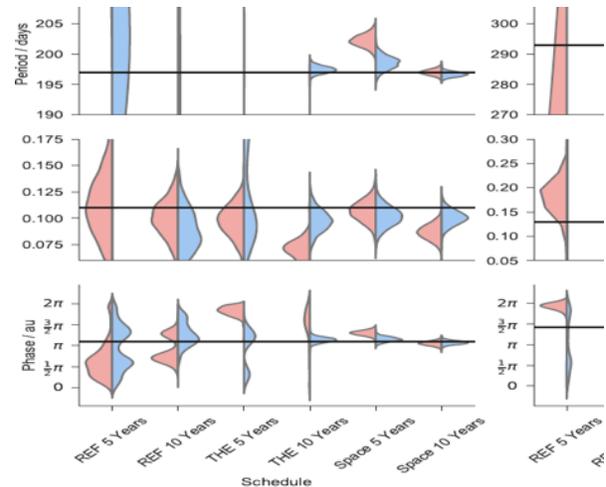

**Figure 1-12**: The violin plots of the posterior distributions for the parameters of the two Earth-like planets in System-2 from Hall et al. (2018). Results for six different observation schedules are plotted here. The distributions colored in blue are the results for data including only Gaussian noise; the red results are when stellar "noise" is also added to the dataset, however corrected for to a level of 75%. The horizontal line in each plot shows the correct parameter value. A narrow, single peaked, Gaussian-like distribution around the correct parameter is the desired result from the analysis; the space-based sampling is optimum for obtaining well constrained solutions around the correct parameters.

Thus, the application of RV color in mitigating stellar activity from visible PRV spectrometers on the ground is currently limited to more active stars. Zechmeister et al. (2018) demonstrated RV color can be used in the red optical to correct activity to 1 m/s.

We explore the impact of this advantage by simulating RV time-series for EarthFinder. We use the same simulations from Section 1.2.3 and add in chromatic stellar activity. We use the StarSIM 2.0 code (Herrero et al. 2016) that models stellar activity from starspots and plages at multiple wavelengths. We simulate an active Sun-like level of stellar activity (~20 m/s peak-to-peak, **Figure 1-13**) for one typical flagship mission direct imaging target: HIP 61317. We simulate N=445 observations in our simulated 5-year PRV EarthFinder survey with 3 cm s[-1] photon noise and 3 cm s[-1] instrumental noise added in quadrature. We choose HIP 61317 for multiple reasons: (1) it is close to the median number of observations for our EarthFinder simulation (N=349); (2) we randomly simulate for this star a 5-planet system analogous to our own Solar System, with a Mercury-like (P=88d, K=2 cm/s), Venus-like (P=204d, K=0.21 m/s), HZ super-Earth (P=307d, K=0.81 m/s), massive Jovian (P=3300d, K=130 m/s), and Neptune-like analogs (51000d, K=0.24 m/s), with moderate e<0.25 eccentricities; and (3) for our ground-based survey simulation, a comparable number of observations are simulated (N=473), much greater than our ground-based median (N=124) and slightly greater than the number of simulated EarthFinder observations. The latter ensures that we are comparing two different time-series with approximately the same number of observations. The chief differences between the space- and ground-based surveys are the cadence, and the additional availability of the determination of the RV color for EarthFinder that is not available at a useful precision from the ground. Our space cadence is uniform random during HIP 61317's visibility windows - it is not in the continuous viewing zone - instead of uniform (daily) sampling as in Hall et al. (2018), and we simulate the EarthFinder precision at 3 cm/s instead of the HARPS3 30 cm/s.

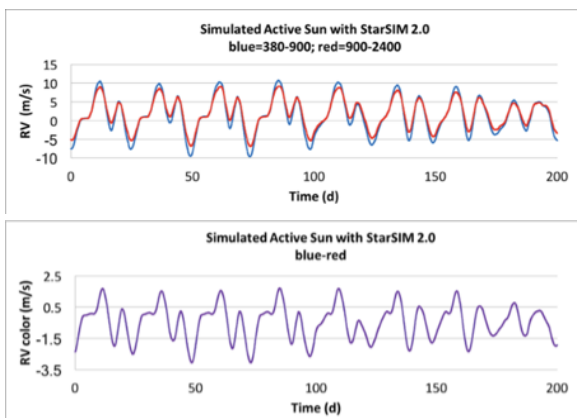

**Figure 1-13**: **Top**: A portion of our simulated stellar activity 5 yr time-series based upon the active Sun for the visible (blue) and NIR (red) arms of EarthFinder, with peak-to-peak changes of ~20 m/s. **Bottom**: The visible minus NIR RV color, showing that the amplitude of the RV color obtained with EarthFinder is large compared to our measurement precision. **To first order, the RV color is proportional to the RV stellar activity in the visible band and completely free of any planetary signals.** Thus, we can exploit this fact to correct the stellar activity in the visible arm. The correspondence is not perfect, due to wavelength-dependent differences in limb-darkening and convective blueshift. Although not accounted for here, such effects can be modeled.

For this analysis, unlike in Section 1.2.3, we do not assume perfect knowledge of the longitude of periastron, and rather we let all of the Keplerian orbital elements be free parameters in our analysis. For both the ground- and space-based cadences, with no correction to the stellar activity, and modeling the stellar activity with a GP, the Bayesian log-likelihoods from RADVEL supports the detection of all but the Mercury-like analog (Fulton et al. 2018). This implies that the detection and masses of HZ Earth-mass analogs can be recovered with GP modeling of stellar activity and sufficient cadence. However, from the ground, only one of the eccentricities is correctly recovered (the massive Jovian analog). The rest of the orbital solutions "blow up" to e=0.5-0.65 (the maximum cutoff in our recovery tests), putting these detections into doubt. **With an incorrect recovered eccentricity for the Super-Earth analog, the orbit information retrieved from the ground is not useful for informing future flagship direct imaging observations.**

We next apply a correction to the stellar activity for our EarthFinder simulation of HIP 61317. We take into account the relatively lower RV information content in the NIR arm compared to the visible, the measurement uncertainties and random sampling. We utilize a simple linear model, assuming the RV color is proportional to the visible stellar activity signal. This is correct to first order, and this is the simplest approach we can take. Unfortunately, we did not have time to model the additional information available from the wavelength dependence of limb-darkening and convective blueshift. In particular, frequency filtering would also greatly benefit our analysis. We also model the RV color with a GP for deriving the GP hyper-parameters and recover the rotation period and spot lifetimes without contamination from the planet signals requiring a joint analysis. This yields important rotational activity phase information that we did not take advantage of in our simple model.

*With this simple linear model, we are able to reduce the activity RV rms by a factor of 62% in the visible channel, which is better than any result that has been achieved by the ground to date with sophisticated line-by-line analysis. Modeling the residual activity again with a GP, we are able to recovery the correct eccentricity and phase of the HZ super-Earth analog, which was not possible in our ground-based simulation.*

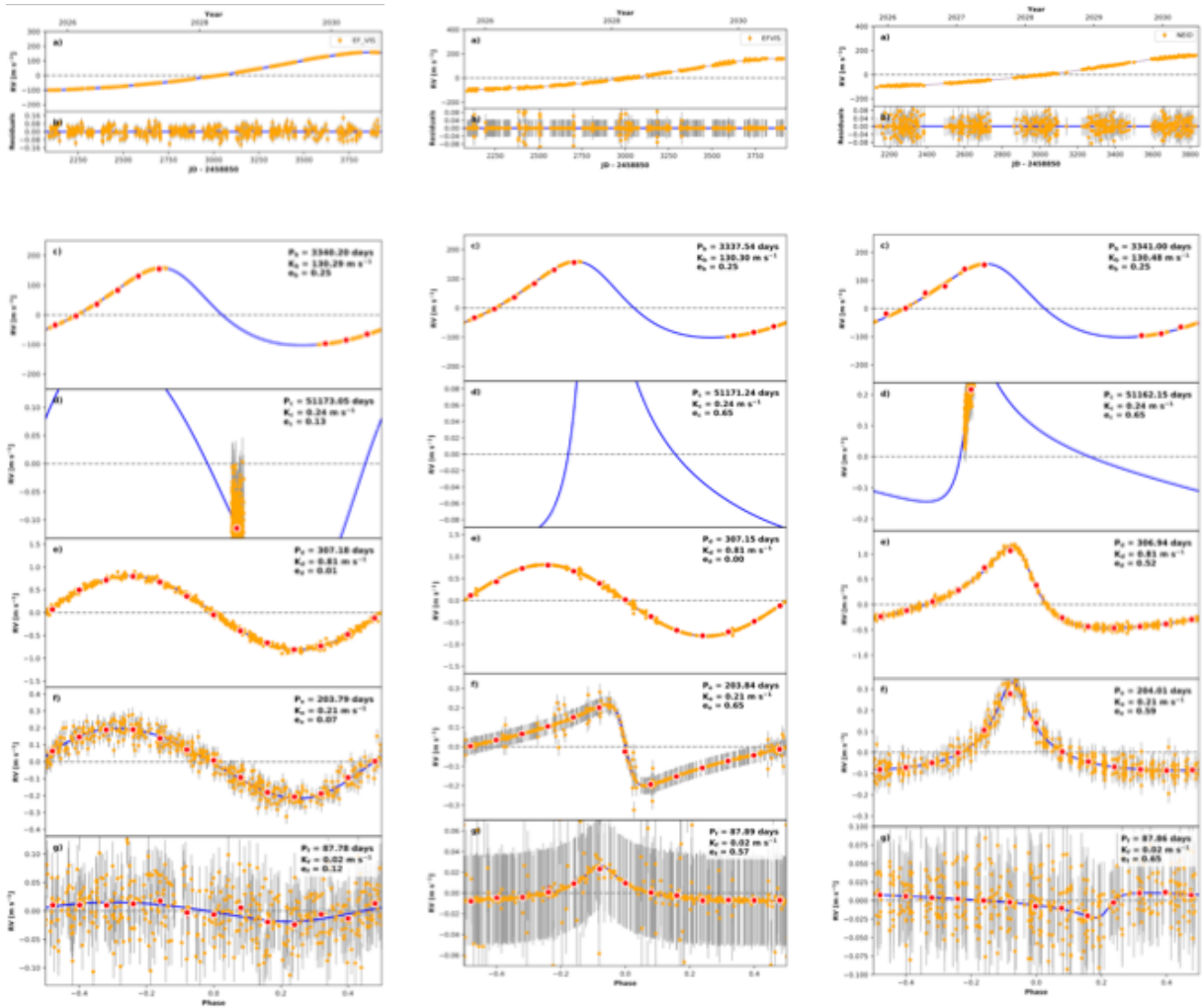

**Figure 1-14**: (**Left**) Our simulated visible EarthFinder RVs for HIP 61317 with perfectly corrected stellar activity. From top to bottom are the full RVs, followed by the best-fitting RV time-series phased for each planet isolated from the other four from RADVEL. With stellar activity perfectly corrected, EarthFinder recovers the detection and orbits of all but the Mercury-like planet on the bottom. (**Center**) Our simulated EarthFinder RVs for HIP 61317 with stellar activity corrected with a simple linear model from the RV color (62% reduction in rms) and a GP model for the residuals. (**Right**) Our simulated ground-based RVs with stellar activity modeled with a GP, but with no initial activity correction. **From space, the orbital information of the HZ super-Earth is recovered, but from the ground it is not and the eccentricity blows up.** In both cases, the Venus analog is recovered, but the eccentricity & phase are not (although EarthFinder better recovers the orbital phase). This is not insurmountable however, as our activity correction was the simplest available to EarthFinder, and we have not yet developed a more sophisticated analysis.

**Figure 1-15** visually shows the superiority of the cadence advantages of EarthFinder in mitigating stellar activity and for planet detection. Without stellar activity, the ground-based RV time-series shows a strong 1-day cadence alias, significant power at periods less than 10 days due to no planets, and a seasonal harmonic. All of these cadence aliases are not present in the EarthFinder RV time-series. Even with multiple longitudinally-spaced telescopes and PRV spectrometers on the ground, the one-day aliasing of each individual site will remain with the zero-point velocity offsets between telescopes that must be modeled.

With the addition of stellar activity, the cadence aliasing at periods of less than 10 days is much more prominent from the ground than in space, the periodogram peaks from the stellar activity are much cleaner for EarthFinder, and the power in the HZ super-Earth analog is greatly reduced from the ground, while still present and statistically recoverable in the EarthFinder RV time-series.

## 1.4.4 HIGH SNR, HIGH RESOLUTION & LINE BY LINE ANALYSIS

Spectral line by spectral line analysis is a critical technique to differentiate between activity signals and real planetary signals. Such analysis allows us to better understand how stellar activity perturbs stellar spectra and therefore RV measurements. In return, this will allow us to strongly mitigate the effect of stellar activity. Stellar activity modifies the shape of spectral lines and therefore the highest resolution is desirable to characterize this perturbing signal better.

A preliminary study from Desort et al. (2007) shows that a resolution higher than R=100,000 allows us to measure a more significant signal for the bisector inverse slope (BIS), a proxy for the asymmetry variation of the cross-correlation function. Similar conclusions were obtained when using the SOAP 2.0 simulation (Dumusque et al. 2014). Additionally, a study from Davis et al. (2017) has also shown that going to higher resolution is key in disentangling planetary from activity signals in RV measurements. Preliminary results show that a mitigation of stellar activity of nearly a factor of two is possible (Dumusque 2018) and there is optimism for much greater improvement, particularly from space. In agreement with Davis et al. (2017), Cegla et al. (2019) also argue that high spectral resolution is necessary if we are to use the line asymmetries as diagnostics of stellar noise from surface magneto-convection (i.e. granulation). This is because the instrument response of a low-resolution spectrograph will act to smooth out the line asymmetries, making it more difficult to use them as stellar noise diagnostics. For example, see **Figure 1-16** from Cegla et al. (2019), which shows the line bisector from disc-integrated Sun-as-a-star model observations of surface magneto-convection for the Fe I 6302 Angstrom line, before and after convolution with various instrumental profiles.

> *EarthFinder will provide the high resolution (R~150,000) required to perform line-by-line studies, which will enable us to mitigate the perturbing effects from stellar activity and therefore allow us to find the signatures of Earth-like planets.*

From our current knowledge of rotationally-modulated stellar activity, two main effects on the RVs can be distinguished: the flux effect that exists because spots and faculae have a different temperature and therefore contrast with respect to the photosphere, and the convection effect that is due to inhibition of the convection by strong magnetic fields (Section 1.4.1). Those two effects will have a different impact on different spectral lines; this is due to different sensitivities to temperature of the elements/transitions at the origin of the spectral line, different sensitivities to magnetic field (Zeeman broadening), but also due to different formation heights of spectral lines in the atmosphere, which implies a different convective blueshift (e.g. Reiners et al. 2016).

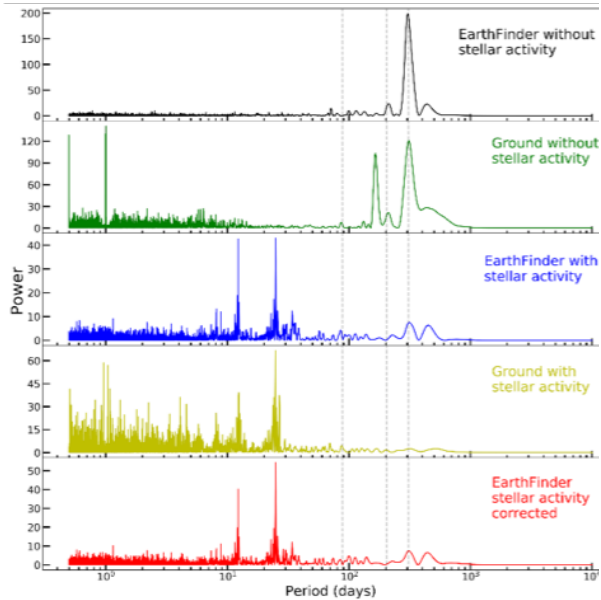

**Figure 1-15**: Bootstrap periodograms of the EarthFinder and ground-based RV time-series simulated for HIP 61317, with the massive Jovian-analog removed. The Mercury, Venus, and HZ super-Earth analog orbital periods are indicated with the vertical dashed lines.

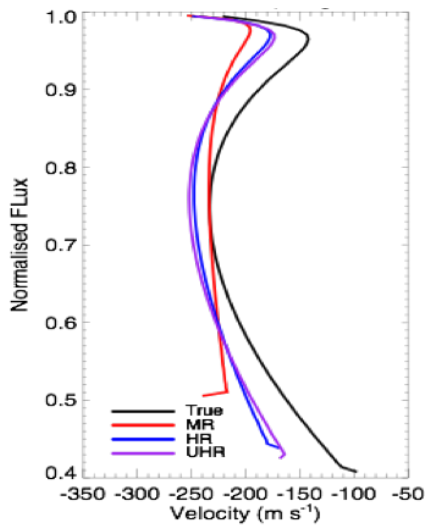

**Figure 1-16**: Line bisector of the average model line profile from Cegla et al. (2019) before (iblack) and after it was convolved with an instrumental profile corresponding to three modes: Medium Resolution (R = 70,000), High Res (R = 140,000), and Ultra-High Res (R = 190,000).

From our current knowledge of rotationally-modulated stellar activity, two main effects on the RVs can be distinguished: the flux effect that exists because spots and faculae have a different temperature and therefore contrast with respect to the photosphere, and the convection effect that is due to inhibition of the convection by strong magnetic fields (Section 1.4.1). Those two effects will have a different impact on different spectral lines; this is due to different sensitivities to temperature of the elements/transitions at the origin of the spectral line, different sensitivities to magnetic field (Zeeman broadening), but also due to different formation heights of spectral lines in the atmosphere, which implies a different convective blueshift (e.g. Reiners et al. 2016).

To distinguish stellar activity effects that will influence each spectral line differently from a planetary signal that affects all spectral lines in the same way, it is crucial to measure the velocity, width and asymmetry variations of each individual spectral line. This requires high-SNR spectra, but also the highest resolution possible. In Davis et al. (2017), the authors found that it is possible with spectral resolution higher than R=100,000 to observe that stellar activity has a different impact on each spectral line. In Thompson et al. (2017) and Wise et al. (2018), the authors look at the variation in core flux and in equivalent width (EW) of each spectral line. They show that both parameters are strongly correlated with stellar activity proxies for certain spectral lines. Preliminary results from Wise et al. (2019) show that line excitation energy is a strong predictor of how activity-sensitive a spectral line is, as can be seen in **Figure 1-16**. Line-by-line analysis is therefore a powerful tool to better understand stellar activity, how it influences stellar spectra, and therefore how it induces spurious RV signals.

Rejecting activity-sensitive lines when calculating the RV could mitigate the impact of stellar activity. However, the core flux and EW can vary without creating a line asymmetry, which is needed to create a spurious RV effect. Therefore, to go a step further, Dumusque (2018) measures the RV of each individual line in stellar spectra and demonstrated that the RV of certain lines is strongly affected by activity, as can be seen in **Figure 1-18**. By only measuring the RV on the spectral lines that are the less affected by stellar activity, it is possible to mitigate its impact by 38%, as can be seen in **Figure 1-19**.

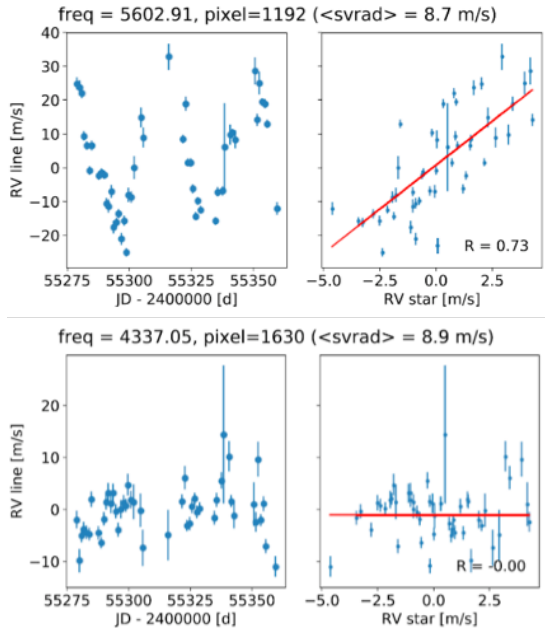

**Figure 1-18**: RV of two spectral lines in the 2010 RV data set of Alpha Centauri B. Each plot is divided in two subplots, on the left, the RV of the line as a function of time, on the right the correlation between the RV of the line and the RV of the star, which for those data is a good proxy for stellar activity. The RV of the line presented on the left is not correlated with stellar activity, while the one on the right is strongly correlated (Dumusque 2018).

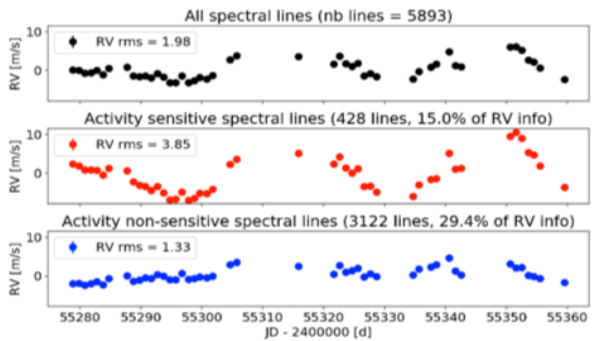

**Figure 1-19**: RV data of Alpha Cen B strongly affected by stellar activity. We show here the RV measured using all the spectral line (top) only the very affected ones (middle) and the less affected ones (bottom). By optimally selecting the lines to measure RV, the bottom selection shows it is

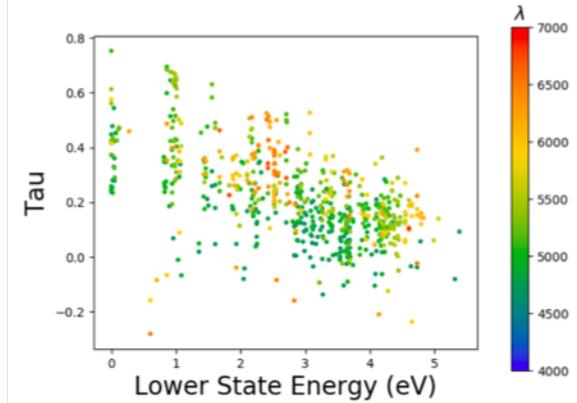

**Figure 1-17**: Kendall's Tau correlation coefficient between line core flux and S-index (Vaughn et al. 1978), versus energy level of the electron configuration required before each absorption can occur. Each point corresponds to a given spectral line color coded according to its wavelength. This previously undiscovered correlation will be exploitable by EarthFinder at higher SNR for mitigating stellar activity.

possible to mitigate stellar activity by 38% (from Dumusque 2018).

## 1.4.5   *CONTINUUM DETERMINATION*

> *The space-based spectra obtained with EarthFinder will not be affected by flux continuum normalization uncertainty which will provide spectral diagnostics that are less noisy, allowing us to much better mitigate stellar signals. In addition, space-based spectroscopy will also allow us to perform spectrophotometry, therefore giving another very important diagnostic for correcting stellar signals.*

One major limiting factor to PRV measurements is from the stellar surface magneto-convection, called granulation (see Section 1.4.1). To overcome this stellar noise, we need to find ways to disentangle it from the Doppler-reflex motion of planetary companions. One approach for this is to study how the stellar absorption lines change in response to surface granulation and how those changes are related to the RVs that we derive.

Recently, Cegla et al (2019) showed that several diagnostics derived from the stellar lines correlate strongly with the convection-induced

RV shifts; thus, we may be able to use the stellar line profile variations as convection-noise mitigation tools. These authors created sun-as-a-star simulations based off a granulation parameterization derived from 3D magneto-hydrodynamic solar surface simulations. In particular, they simulated the Fe I 6302 angstrom line from MHD simulations with an average magnetic field of 200 G (i.e. a bit more magnetic relative to the quiet Solar photosphere). As such, their model observations represent a star with only convection as a contributing stellar noise source and may underestimate the convection noise by a factor of 3-4 due to the increased magnetic field (which suppresses convective motions). Nonetheless, these model observations still offer a window, beyond instrumental limitations, into the nature of the stellar surface convection noise.

One of the strongest convection noise diagnostics came from measuring the variations in the stellar line profile depths/contrasts. The physical driver behind this is likely due to the fact that convective granules are formed higher in the photosphere and therefore have deeper line depths. Consequently, if more granules are present on the star at a given time we expect the disc-integrated profile to be both deeper and more blueshifted. However, it is possible that some combinations of granules to intergranular lane components could potentially produce the same ratio from continuum to line core; if this were the case, then these degeneracies would mean information is lost and correlation with RVs is degraded when continuum normalizing ground-based spectra.

In line with this, the absolute line depths would only be available from a spectrometer in space, as ground-based data must be continuum normalized to remove contamination from Earth's variable atmosphere. The simulations in Cegla et al. (2019) show a strong correlation between the stellar line depth and the convection-induced RV shifts. Moreover, we also see that continuum normalizing the stellar lines does indeed increase the scatter in this correlation; in fact, this degradation in the correlation is sufficient to completely negate this diagnostic's noise mitigation ability (Cegla et al. 2019). The total variations in both line depth and RV will be greater if the magnetic field is lower, e.g. if it were closer to the quiet Sun. Nonetheless, the requirement to continuum normalize means the line depth will be a less powerful noise diagnostic from the ground, and may not allow for the correction of any of the convection induced variations without going to space with EarthFinder.

The advantages of not needing continuum normalizations in space enhances our ability to detect and precisely measure stellar signals, making a space-based high-precision spectrometer, such as EarthFinder, ideally suited to aid in the stellar noise mitigation in ways not achievable by ground-based spectrographs.

> *An important aspect in measuring precisely spectral line variations from spectrum to spectrum in ground-based observations is getting an excellent continuum normalization. Even if done carefully, a residual noise will always affect our measurements. Spectra obtained with EarthFinder will not have this added noise from continuum fitting.*

Additionally, a space-based spectrometer would also mean we could integrate the absolute flux underneath the spectra; this would act as a direct proxy for the photometric brightness variations -- without the need for an additional instrument. This is also significant because it adds another type of stellar noise diagnostic not achievable from the ground.

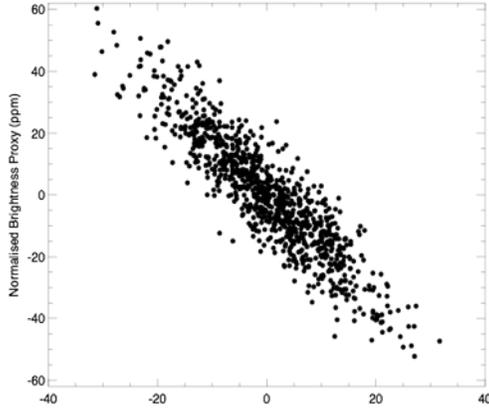

**Figure 1-20**: A brightness proxy versus RV, where the proxy is determined by the integrated area underneath the disc-integrated model line profiles. This area has been normalized by its maximum value, mean-subtracted, and converted to parts per million (Cegla et al. 2019).

For example, Cegla et al. (2019) has shown the convective-induced brightness measurements, as derived from integrating the area under their simulated line profile, are even more strongly correlated with the convective-induced RV shifts and may allow us to remove >50% of the RV variability; see **Figure 1-20**. Moreover, convective-induced brightness variations would require precision measurements of 10s of parts per million (ppm), which is only achievable from space. Such space-based photometric missions will be difficult to coordinate with ground-based RV follow-up. However, a space-based spectrometer, like EarthFinder, would naturally provide simultaneous RV and brightness proxy measurements. Hence, EarthFinder offers a variety of unique stellar noise mitigation advantages over other high precision, ground-based spectrographs.

## 1.5  GENERAL ASTROPHYSICS WITH EARTHFINDER?

### 1.5.1  *INSTRUMENT CAPABILITIES*

> *Beyond the primary science case of PRVs, EarthFinder offers an unprecedented platform for space spectroscopy for a number of unique science cases.*

In this section we highlight a number of secondary science programs enabled by EarthFinder. We plot the simulated sensitivity of EarthFinder in **Figure 1-21**. For the visible arm, the exposure time for a Teff=5700 K object is, to 1% over V=5—20 mag:

$$t_{exp} = 4153.5 \ sec \left(\frac{SNR}{10^4}\right)^2 10^{m_V/2.5}$$

With mirror coatings optimized for visible and NIR transmission, the UV arm of EarthFinder has reduced transmission (~20%, requiring longer integration times than a UV-optimized telescope).

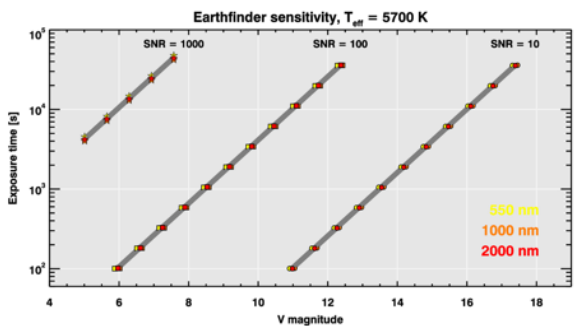
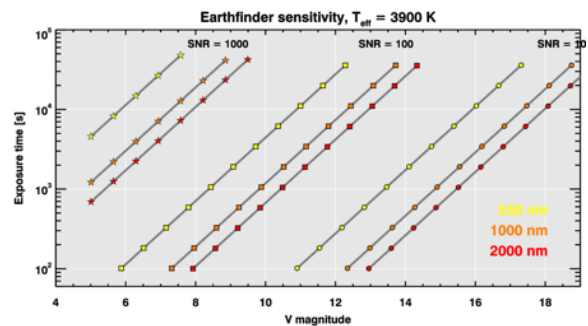

**Figure 1-21**: We plot the integration times required for R=150,000 to reach the labeled SNR at the indicated wavelengths as a function of apparent magnitude for a Sun-like star (left) and M dwarf (right). Fainter targets can be reached in tractable integration times at degraded spectral resolution from spectral pixel binning, which is more cost effective than hardware for lower resolution modes without a significant detector noise penalty.

## 1.5.2   *STELLAR ASTROPHYSICS*

**Stellar Oscillations** | In recent years, Kepler photometry has been used to detect the intensity variations caused by stellar pulsations modes of FGK stars. Seismology provides a direct measure of the stellar density and can be used to derive the stellar radius, mass, and age of the stars as well as yielding a better understanding of the internal structure of the stars. The excitation of solar-like oscillations is not fully understood. While it is clear that the modes are excited by convection, our inability to model convection properly translates to our inability to explain the excitation of these modes. Consequently, only the frequencies of modes are used in asteroseismic analyses; mode widths and amplitudes are ignored.

Helioseismologists have long argued that seismic observations in multiple spectral ranges can provide additional information (see e.g., Hill 2009, Salabert et al. 2009). Howe et al (2011) spatially averaged images taken by the AIA instrument on SDO and showed that some spectral ranges have lower granulation noise than others, which is an indication that **granulation noise and its effects are highly wavelength dependent.** Simultaneous observations in different wavelengths will also give a unique window for studying wave propagation in stellar atmospheres (Finsterle et al. 2004) that can help characterize the acoustic-cutoff frequencies and test the accuracy of one of the key global asteroseismic scaling relations, that $\nu_{max}$, the frequency of maximum acoustic power, is proportional to the acoustic cutoff frequency. This feature is routinely used to determine stellar masses and radii from asteroseismology.

Radial velocity variations from pulsations of giant (km/s, weeks to months), sub-giant and some dwarf stars (a few m/s, a few minutes), are within the reach of EarthFinder. Higher-order (octopole) pulsation modes are accessible with spectroscopy from line shapes at high-resolution and are inaccessible with photometry, particularly for hot stars that were not as well characterized with Kepler. EarthFinder would offer cadence advantages that are not possible from the ground at a single observatory.

**Stellar Ages** | A seismic analysis of stellar cores is the only way to determine precise ages of stars. Kepler has provided asteroseismic data of a few tens of exoplanet hosts and TESS will do the same. However, both Kepler and TESS observe in intensity. It is known from solar data that power spectra of intensity data have increased granulation noise in the low-frequency regimes making it impossible to determine the low-frequency modes that are most sensitive to the innermost regions of a star (see **Figure 1-22**). Also, different spectral lines are produced at different heights in the atmosphere, and hence mode-excitation related properties (amplitude and widths) should be different at different spectral ranges. EarthFinder will allow us to determine the best spectral regions to use for seismic analyses of exoplanet host stars, how the information from different spectral regions may be combined to give the best signal-to-noise in the oscillations, and how long we need to observe a certain star to precisely infer its properties.

**Stellar Abundances** | The 0.9-2.5 μm band contains features of water, methane plus numerous other species. EarthFinder will be able to create an atlas of spectra of a variety of stars with high sensitivity and spectral resolution inaccessible from the ground at the precision obtainable from space due to tellurics. Water vapor is a dominant opacity source for M dwarf atmospheres (early and late), and along with other molecular species limits our ability to produce synthetic spectra that accurately match observed NIR spectra and colors of M dwarfs (e.g., Passegger et al. 2016, Auman 1967, Langhoff et al. 1996, Alexander et al. 1989). EarthFinder will enable us to obtain accurate and

detailed high-resolution measurements of the hot water and other molecular species opacities in the NIR to further improve upon existing model atmosphere codes (e.g. BT-SETTL models, Baraffe et al. 2015, Allard et al. 2014).

### 1.5.3 *EXOPLANET AND DISK SCIENCE APPLICATIONS*

**Exoplanet Atmospheric Loss** | HST is currently the only facility capable of carrying out UV spectroscopy giving EarthFinder a unique role in the post-HST era. EarthFinder will offer continuous coverage of exoplanetary transits by avoiding Earth occultations, SAA interruptions, geocoronal contamination, and the breathing effect that limit HST observations. EarthFinder will also yield simultaneous observations of exoplanet atmospheres from near-UV to near-IR, e.g. He I 10830, which probe the entire atmospheric structure in a single transit. This is essential given the temporal variability observed in exoplanet upper atmospheres (e.g. Lecavelier et al. 2012).

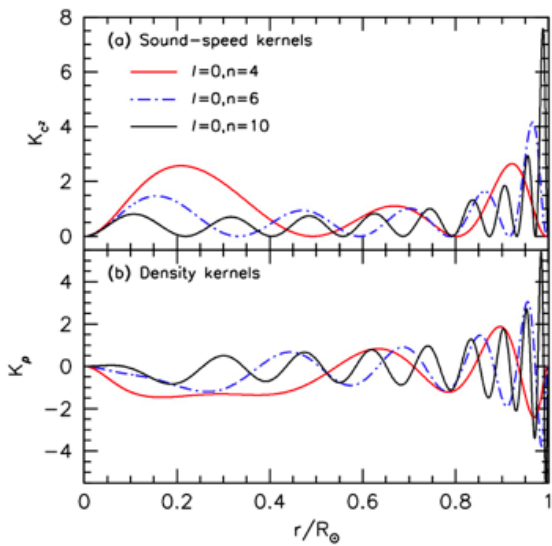

**Figure 1-22**: Sensitivity kernels for (a) sound-speed and (b) density for a few radial modes. Note that the lower frequency model (n=4) has the largest sensitivity at the core. Such low-frequency modes are lost in granulation noise in brightness intensity measurements.

Massive atmospheric escape has been detected from several Jupiter and Neptune-size exoplanets in the Lyman-alpha line of neutral hydrogen (e.g. Vidal-Madjar et al. 2003, Ehrenreich et al. 2015). Because of their spatial extension the resulting expanding atmospheres yield deep transit signatures, making them much easier to probe than the lower atmosphere. Observations in the FUV stellar Lyman-alpha line probe the outermost atmospheric layers but are limited to the closest systems by interstellar medium absorption (eg Bourrier et al. 2017). There is thus a strong interest in searching for transit signatures in the lines of atoms and ions lifted in the upper atmosphere. Many of these species have strong transitions in the near-UV range (280-380 nm), such as MgI and MgII, Mn I, Ti I, Ti II, Al I, as well as many excited lines (e.g., Fe I, Fe II, Ti I, and Ti II).

With high spectral resolution in both near-UV/IR ranges, EarthFinder will be a unique tool to detect and characterize the upper atmospheric layers of transiting and non-transiting close-in planets. Escaping exospheric layers interact with the wind and radiation from the star, and form extended tails that yield broad, blueshifted absorption signatures. Resolving finely their spectral absorption profile will bring strong constraints on both the properties of the stellar environment and that of the atmospheric outflow (e.g. Haswell et al. 2012, Bourrier et al. 2014, 2016). EarthFinder will allow for the study of a large sample of planets, which is essential to determine the impact of atmospheric mass loss on the exoplanet population (e.g. Ehrenreich & Desert 2011, Owen & Jackson 2012). Closer to the planet, species in the thermosphere still yield deep but narrow absorption lines that shift with the planet motion. Resolving the spectral absorption profiles of different species in the expanding thermosphere will inform on its chemical composition, ionization degree, as well as its temperature, density, and dynamical structure (eg Yan & Henning 2018, Cauley et al. 2018). Post HST, EarthFinder will also be uniquely suited to the nascent field of disintegrating ultra-short-period planets (< 1 days). These small rocky planets could be the remnant cores of larger progenitors that lost their gaseous atmosphere, making them invaluable probes of planet interior structure and past evolution. Measuring metals sublimating and escaping from this envelope would bring direct constraints on the erosion rates of these planets and the composition of planetary cores.

**Exoplanet Transmission Spectroscopy** | EarthFinder will complement JWST and ARIEL/CASE by providing visible and near-IR coverage at high spectral resolution (R~10,000-50,000 by binning from R~150,000) observations for transiting hot Jupiters orbiting FGK dwarfs and Neptunes orbiting M dwarfs with H<6 mag which might be discovered by the TESS mission.

**Protoplanetary Disks** | One of the most intriguing questions in the study of terrestrial planet formation and a focus of NASA's astrophysics plan, is how water and organics are transported to their surfaces, and whether or not water is a common ingredient during their early formation and evolution. We can place our Solar System in context by studying the astrochemistry of nearby young stars and their circumstellar disks. EarthFinder will be able to provide high resolution spectroscopy unobstructed by the Earth's atmosphere of the volatile content in nearby bright young stars. Targets include Herbig AE stars and FU Ori outburst stars such as the most famous examples of V2364 Ori, FU Ori, V883 Ori, HD 142527, Elias 2-27, HL Tau, TW Hya, AB Aur and HD 163296 with NIR magnitudes from 6 to 9 mag.

We are interested in following the major molecular carriers of volatiles including CO, and H2O as well as the carriers of organics (CH4, and potentially HCN and C2H2), and thus we require high spectral resolution for tracing the kinematics of lines. This has been done using ALMA in recent work by Teague et al. (2018)

and Pinte et al. (2018) at large distances from the central star, but could be uniquely applied in the <10 AU region with EarthFinder.

### 1.5.4   EXTRAGALACTIC SCIENCE

The UV capabilities of EarthFinder will be of critical importance in the post-HST era and before the launch of one of the flagship missions now under study, LUVOIR or HABEX, which are likely to have UV spectroscopic capabilities. EarthFinder would be able to spectroscopically survey bright, nearby galaxies.

For example, EarthFinder will be able to simultaneously monitor wavelengths ranging from ~280 nm to 2500 nm. This would enable simultaneous Reverberation Mapping (RM) of Hβ (468.1 nm), and Mg II (279.6 nm), which has never been done before at high precision, and would allow us to address many questions related to the utility of Mg II in RM studies. Having access to such a broad wavelength range spanning the NUV to the IR would allow us to do simultaneous RM of numerous emission lines and a wide variety of continuum windows and to explore the connections between the NUV/Optical/IR continuum.

# 2   ENGINEERING, INSTRUMENT, & MISSION DESCRIPTION

## 2.1   OVERVIEW

In order to achieve its goals of discovery and mass measurement of rocky planets at habitable-zone separations around sun-like stars, the EarthFinder Probe requires exquisite understanding and control of both instrumental and astrophysical noise sources. The instrument study presented herein preserves the well-founded principles of Doppler spectroscopy as practiced from the ground, but reconsiders the approach in a space-based context. The environment from space simultaneously allows important advantages, such as the lack of telluric contaminants (Section 1.3), continuous viewing and high observing cadence (Section 1.4.2), the ability to use compact high-throughput optical systems in a small, relatively lightweight payload with high thermal-mechanical stability. These crucial advantages are not available from the ground.

Instrument requirements and notional designs are derived from the EarthFinder mission's science objectives. The measurement principle is based on the ultra-precise centroiding of stellar absorption lines in search of small, periodic signals due to Doppler shifts. EarthFinder's targets are all bright stars in the solar neighborhood, and the probe's ability to detect and measure small planets rests on four overarching factors: 1) the photon collecting capability, which when combined with the stellar spectral-line density, determines the ultimate measurement precision; 2) the instrument systematic errors on short-to-mid-timescales, which degrade single measurement precision; 3) the long term (multiple years) measurement stability; and 4) the target observing cadence, which determines the ability to track, model and remove stellar activity noise, so as to reveal the presence or absence of underlying Doppler signals.

This engineering concept has leveraged on-going developments in technologies as well as improved understanding of velocity error sources, both aimed at pushing the measurement boundaries of assorted next-generation ground-based spectrographs; it assimilates the latest developments from a new generation of seeing-limited spectrographs such as NN-EXPLORE Exoplanet Investigations with Doppler Spectroscopy (NEID) (Schwab et al. 2016; Halverson et al. 2016), as well as the adaptive-optics fed, compact, diffraction-limited spectrographs such as Palomar Radial Velocity Instrument (PARVI), iLocator and Keck- High Resolution Infrared Spectrograph (HISPEC) (Crass et al. 2018). These diffraction-limited spectrographs, together with a diffraction-limited telescope and optical system, are exceptionally well suited for implementation in a space-based PRV mission (see Section 2.2.3.5). EarthFinder will use laser-frequency comb metrology for calibration. Planned advances in the miniaturizing of self-referenced laser-frequency combs will enable the <1 cms$^{-1}$ level long-term stability needed to calibrate radial velocity observations over a mission lifetime (see Section 2.3).

### 2.1.1   THE TOP-LEVEL REQUIREMENTS

The payload consists of a telescope, distribution fore-optics, and three instrument spectrographs, one each covering the near ultraviolet, optical, and near infrared wavelength bands; for the remainder of the document these spectrographs are simply dubbed as the Ultra-Violet Spectrograph (UVS; spanning 200-380 nm), the OPtical Spectrograph (OPS; spanning 380-950 nm), and the Near-InfraRed Spectrograph (NIRS; spanning ~950-2500 nm). The PRV spectrographs, OPS and NIRS, are fed using single-mode (SM-) fiber feeds, and have very high resolving powers of R~150,000 each to enable line-by-line analyses. They are stabilized by virtue of design, and remaining instabilities are monitored by combs, which produce thousands of individual lines at ~8 GHz separation, each frequency-stable at the <1 cms-1 level over years. UVS has a nominal resolving power of R~3000, and monitors the established stellar activity indicators. All three spectrographs are configured to operate simultaneously.

**Table 2-1:** The Science Traceability Matrix is used to derive the EarthFinder Mission and Instrument functional requirements.

| ç | Science Objectives | Scientific Measurement Requirements | | Instrument Functional Requirements | Projected Performance | Mission Functional Requirements (Top Level) |
|---|---|---|---|---|---|---|
| | | Physical Parameters | Observables | | | |
| Goal 1: Seek out new worlds and determine if they might be habitable | O1: Determine if small (0.8-1.7 R_E) planets exist around nearby Sun-like stars and continuously orbit within the HZ; Survey a sample of FGK stars to reach HZ completeness > 75% for exo-Earths ($m\sin i = 0.5 - 4.3$ M_E at $i = 60$ deg). | Periodic changes and trends in the radial (line-of-sight) velocity of the star to determine semi-major axes, eccentricities, and minimum masses of planets to 10% for 1 Earth-mass planets at $i = 90$ deg. Use orbital elements and properties of the star to infer the effective temperature and potential for habitability of the planets. | Stellar Spectrum: Measure line centroids relative to a local wavelength standard with noise equivalent of < 10 cm/s (per epoch) Stellar Activity: (a) Spectral lines shapes over a broad wavelength span (b) Equivalent widths of activity indicator lines to 1% (c) Spectrophotometry to < 1% at low resolving power R=100 | Spectral range: 0.4-2.4 µm with median resolving power R > 140,000 Instrument Doppler noise equivalent < 10 cm/s (in 1 hr) and 1 cm/s (over mission duration) UV spectral range: 0.24-0.4 um with resolving power R > 1000 Photometer | Two stabilized Echelle spectrometers cover 0.4-2.5 µm range simultaneously with R = 150,000 Spectrograph Doppler noise ~5 cm/s/hr[1/2] UV grating spectrometer with R=3000 0.5 % relative spectrophotometry | Observe 60 stars 80 times a year during viewing period(s) Time baseline > 4 yr Telescope aperture > 1.2 m, diffraction limited at 0.4 µm. Pointing = 10 mas (1-σ 2 axis jitter) Spacecraft velocity < 1 cm/s FOR = 71% of celestial sphere |
| | O2: Survey a nearby sample of sunlike stars and cool dwarfs (1.1 M_S ≥ M ≤ 0.1 M_S) and determine the architecture of their planetary systems out to beyond snow-line orbits P_ORB ≤ 5 yr. | | Stellar Spectrum: Measure line centroids relative to a local wavelength standard with noise equivalent of 10-30 cm/s (per epoch) Stellar Activity: (see above) | Same as above | | Survey time > 4 yr |
| | O3: Determine the architecture of young planetary systems (age ≤ 1 Gyr). Survey a sample of young stars and determine the architecture of their planetary systems out to orbital periods of P_ORB ≤ 5 yr. | | Stellar Spectrum: Measure line centroids relative to a local wavelength standard with noise equivalent of 100-500 cm/s (per epoch) Stellar Activity: (see above) | Same as above | | Survey time > 0.5 yr |

The Science Traceability Matrix (STM) (**Table 2-1**) lists the top-level functional requirements for the mission and payload. Source "photon sufficiency" determines the effective collecting area of the EarthFinder telescope, including its optical and detector efficiencies. The operating conditions for the sensors and the need to limit "noisy" background photons determine the operating temperatures of the spectrographs (these are 170 K for UVS and OPS, and 60 K for NIRS). The single mode entendue allows the telescope and the fore-optical system to be operated at room temperature. These are standard requirements and easily met within a wide design space.

The required Doppler precision places strict stability requirements on the spectrographs. As mentioned above, the PRV spectrometers employ SM-fiber feeds. This modal isolation offers two clear advantages relative to multi-mode or seeing-limited instruments:

1. Diffraction limited spectrographs are inherently small in size. The spectrograph size scales with the etendue, which is independent of the telescope size at the diffraction limit. For an EarthFinder-like telescope in 0.5 arcsecond seeing conditions, the spectrograph dimension would increase by a factor of 5. This factor in linear size results in a space-based spectrometer that is between 25 to a 125 times smaller in mass and volume than a seeing-limited one with comparable spectral resolving power. Overall thermal-mechanical stability is easier to maintain within a compact package. Finally, efficient diffraction-limited spectroscopy is only possible across the entire optical band from space.

2. SM- operation completely isolates the spectrographs' optical performance and PSF (and equivalently the Line-Spread-Function (LSF)) from upstream telescope aberrations and residual spacecraft pointing error. This decoupling prevents upstream errors from degrading the downstream spectrograph Doppler performance.

Reliable coupling of starlight into the fibers and minimization of intensity losses, however, necessitates fine pointing control. To achieve the required pointing, the EarthFinder spacecraft (S/C) attitude control system works together with a payload fine guidance system (FGS) (performance: pointing stability of 15 mas rms 2-axis over a visit, determined to maintain dependable contact (loss no greater than 20% at the optical fiber feed, at the shortest PRV wavelengths ~400 nm). The FGS uses a Charge-coupled Device (CCD) based fine guidance camera (FGC) to monitor the target star motion by using a fraction (0.1) of the light of the on-axis stellar target in the 380-950 nm pass-band and a fine steering mirror (FSM) to compensate for line-of-sight motion. At 100 Hz sampling rates, the FGS meets the noise equivalent performance for targets as faint as V=15, providing the needed pointing performance for all primary and ancillary science cases. To tie the location of the entrance fibers to the FGC, the PRV spectrometers uses back-illuminated reference beacons on the FGC sensor. The static coupling to the fibers is maximized by beam shaping fore-optics (e.g. Jovanovic et al. 2017), which reshape the telescope beam into a quasi-Gaussian pattern. This overcomes the usual ~80% maximum modal coupling efficiency for injecting telescope light into SM-fibers (Shaklan & Roddier 1996). The telescope/fore-optical system may have to rely on a common focus adjusting mechanism, possibly located at the powered mirror M3; given the inherent stability in orbit, EarthFinder relies on one-time focus tuning after launch.

EarthFinder design must minimize un-calibratable thermal-mechanical distortion of the RV spectrographs. The design minimizes temporal distortion by virtue of the S/C orbit, by shielding the instrumentation from extraneous thermal forcing, and by building the spectrograph optics within Silicon Carbide (SiC) glass sandwiches. Analysis shows the need to maintain temperature stability to ~10 mK rms during observations (Section 3). This level of stability on the base temperature and any local gradients can be met with adequate design margins. Better temperature control (~ 1 mK rms), however, is required for the spectrograph focal plane detectors to limit non-ideal detector behavior.

The Doppler velocity error budgets in **Table 2-2** and **Table 2-3** show that the thermal and opto-mechanical engineering will allow the spectrographs to be velocity stable to ~15 cm s$^{-1}$ over the course of a single observation. Any remaining instrument instability must be measured and calibrated out using the EarthFinder comb metrology, for which the mission will rely on a comb-to-starlight common mode factor of 90% and an intrinsic stability of the comb teeth to better than a cm s$^{-1}$. We find that post calibration non-ideal detector behavior (on both the CCD and IR focal planes), which is unmonitored by metrology, is largest single source of instrumental RV systematic. The sources of sensor imperfections that bias RV measurements are the same as the ones that plague other astrophysical precision measurement (such as galaxy shape measurement with Euclid and the Wide Field Infrared Survey Telescope (WFIRST); Shapiro et al 2017) and preflight characterization and in-flight calibration will be required to mitigate them.

## 2.2   SPACECRAFT

Based on the mission's needs for extensive sky coverage, continuous target visibility, uninterrupted integration times on targets of greater than an hour, and the thermal-mechanical stability needs of its instruments, EarthFinder will be placed in an Earth-Sun L2 libration point orbit, or perhaps a heliocentric drift away orbit. In its orbit the spacecraft velocity will need to be known or modeled to better about a cm s$^{-1}$ relative to the solar system barycenter. For this, relatively mature techniques exist for determining the spacecraft velocity vector to an accuracy of a few mm s$^{-1}$ (Hirabayashi et al. 2000, Scheeres et al. 2001).

The point of departure design concept for the EarthFinder is shown in **Figure 2-1**. The payload, including the telescope, is mounted to the spacecraft bus on an athermal interface using a bipod arrangement. A cylindrical baffle surrounds the telescope. EarthFinder has thermal shields affixed to the S/C on the Sun-oriented-side of the telescope baffle, along with solar arrays (also fixed to the S/C) installed outward from the thermal shields. The arrays work together with the thermal shields to isolate the payload from solar insolation and reduce the variability of the heat load due to changes in the S/C pointing. There is a deployable high gain antenna articulated to point towards the Earth. The EarthFinder field-of-regard (FOR) is set by the following:

1.  The range of S/C pitch angles controlling the telescope line-of-sight direction relative to the Sun direction;
2.  The 360-degree rotation range about the Sun line; and
3.  S/C roll keeping the solar arrays towards the Sun.
    As a result of these design parameters, all pointing angles greater than 45 degrees from the ecliptic plane are available for year-round observing. This telescope field-of-regard covers 71 % of the celestial sphere, which means that a comparable fraction of EarthFinder targets will be available at all times. Targets distributed at ecliptic latitudes of less than 45 deg. have two observing seasons a year each lasting between 90 and 180 days, the duration of which depends the target's ecliptic latitude.

The angle of the cylindrical wedge on the baffle determines the pitch angle away from the Sun that prevents direct illumination of the interior. The fixed solar arrays are a compromise. These arrays have a maximum solar flux for a pointing angle lying in a plane perpendicular to the direction of the Sun, and out of plane the flux is reduced as the cosine of the angle, to about 0.71 times the peak power generation capacity. The payload will use about 0.8 kW power at current best estimate, with the cryo-coolers and the optical frequency comb electronics being the two largest consumers.

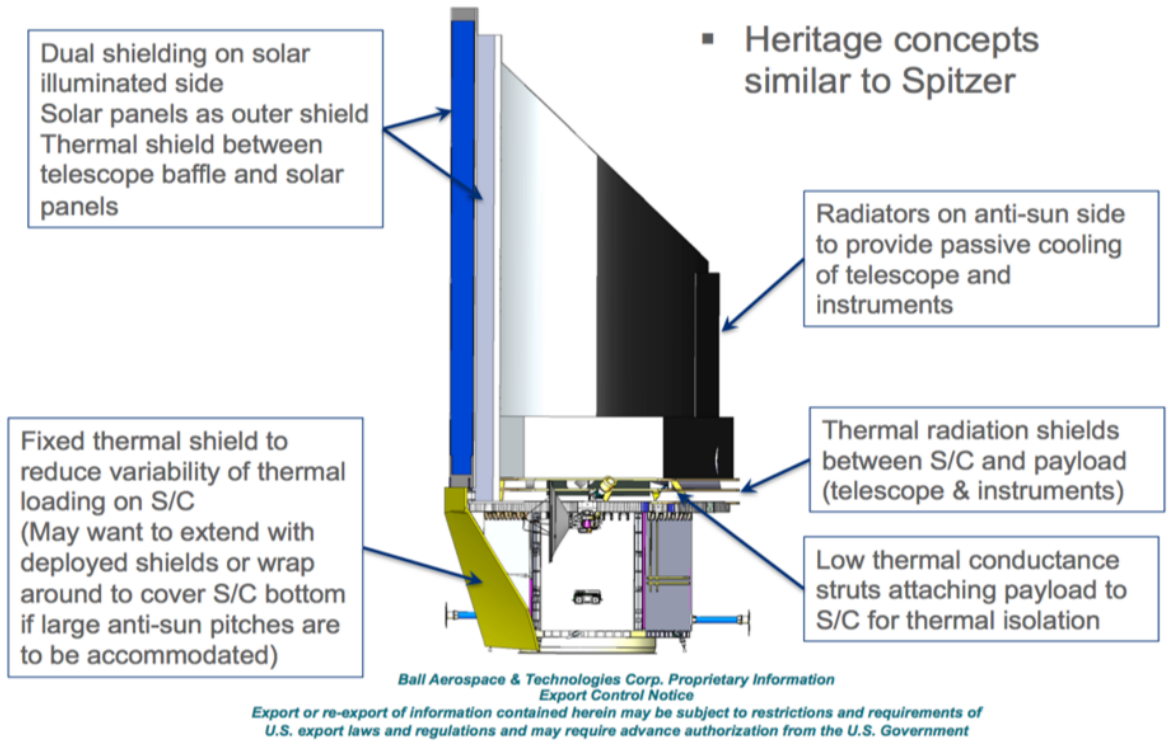

**Figure 2-1:** Thermal management and control of the spacecraft derives much of its heritage from features implemented previously in NASA's Spitzer mission.

The thermal control features for EarthFinder are annotated in **Figure 2-1**. This is the current point design for the thermal management and control architecture. Measures to increase the telescope FOR if desired, include replacing the fixed solar panels with deployable paddles that may be articulated to compensate for the varying pitch of the spacecraft. This extended design might require additional baffling of the telescope and the addition of thermal shielding on the bottom of the S/C to mitigate the solar heating at large anti-sun pitch angles. Such a design would add the complexity of deployable solar arrays and carry additional mass.

### 2.2.1 SCIENCE OBSERVING PROFILE

EarthFinder's primary targets are well known nearby stars. The number of targets, target dwell times, and the number of visits has been determined via a mission simulation assuming a 5-yr nominal duration (see Section 1.2.3). Science observations are conducted in a single operating mode using established methods, allowing for rather straightforward mission scheduling. The orbit, operations and distribution of target stars enable high duty cycle science observations, while conforming to usual solar/lunar pointing angle constraints. Mean target dwells last for 78 minutes and are followed by slews to the next target, separated on the sky by ~20-50 degrees. An observation consists of target acquisition, followed by individual spectral exposures lasting a predetermined total exposure and frame times depending on target brightness, cosmic ray hit rates, etc. With long dwell times per target, and standard slew, settle, and acquisition overheads, EarthFinder will have high observing efficiency.

The observation scheduling will be built on four primary components: target viewing opportunities, spacecraft operational constraints, and putative orbital-period sampling requirements. In creating a mission observing schedule that fulfills science objectives, target viewing opportunities are determined by JPL-developed orbit propagation algorithms and target tracking simulations based on Analytical Graphics, Inc's (AGI) Satellite Toolkit software. Results from these analyses provide all target viewing windows, which account for the various constraints over the course of the mission. Based on the schedule constraints from the current target list and the assumption of 62 targets in the Baseline Mission, EarthFinder could have a schedule margin of ~20 %. The target schedule margin is the possible total number of visits possible in the baseline mission and the total number required to accomplish the prime mission.

## 2.2.2 DOPPLER SPECTROGRAPH REQUIREMENTS

The SM- fibers decouple the telescope and other upstream optics from the PRV spectrographs; any upstream optical instabilities lead to coupling losses and therefore increased photon noise contributions, but no systematic PRV errors. Systematics arise only within the spectrographs and their internal precision requirements are met by: (1) stabilizing the shape of the color-dependent spectrograph PSF, and derived LSF; (2) stabilizing the location of the PSFs on the detector plane; (3) providing a strict wavelength reference standard using optical frequency combs. Allocations for various spectrograph errors are listed in

OPS/NIRS error budgets presented separately in **Table 2-2** and **Table 2-3**.

Thermal and mechanical stability is critical in mitigating any temporal drift and subtle PSF shape changes. Both the orbit and the sun-shield reduce temporal thermal variation and spatial gradients in the spectrograph package. The construction of spectrographs within athermal sandwiches of SiC to reduce overall sensitivity to remaining thermal forcing and gradients (see Section 3). Finally, the spectrometers are built so that there is no need for internal moving parts.

The FWHM of the near Gaussian PSFs on the detectors is chosen to be 3.5 pixels wide (median) in each spectrometer to ensure better than Nyquist sampling of instrument line function and to minimize the size of systematics arising due to interaction with second-order detector imperfections. In addition, the laser frequency comb metrology provides full, in-situ instantiations of the color-dependent PSF across the spectrograph range, which will be put to use in PRV analysis.

The optical frequency combs serve as the natural calibrators for diffraction limited, fiber-fed spectrographs. The combs are EarthFinder's components with the lowest technology readiness. However, the field of optical frequency combs is rapidly advancing and EarthFinder stands to exploit this. For example, research into chip-based micro-resonator combs (Suh et al. 2018) can provide compact, low-power, and highly stable frequency standards at optical and NIR wavelengths. Line spacings of 5-20 GHz, at the natural separations of Doppler spectrographs, are available without the need for any complex modal filtering. Using an octave span in wavelength and f-2f locking, these combs can be operated in a self-referenced mode, giving them < 1 cm s$^{-1}$ long-term stability by tying them to a fundamental SI time standard.

### 2.2.3  *PAYLOAD DETAILS*

The EarthFinder payload consists of a 1.45 m telescope in a Ritchey-Chretien configuration, with Zerodur primary and secondary mirrors in an F/10 configuration, feeding an instrument suite. The instrumentation layout consists of the optical and near infrared Echelles (OPS and NIRS, respectively), the UV grating spectrometer (UVS), the FGS, and support electronics to operate and control these instruments (**Figure 2-2**). The FGC observes the on-axis target to stabilize it to a fraction of the diffraction width on the spectrograph entrance fibers. The telescope and instrumentation bench sit in an enclosure for thermal control and stability. The total mass of the payload is 230 kg at current best estimate with 30% contingency (**Table 2-6**).

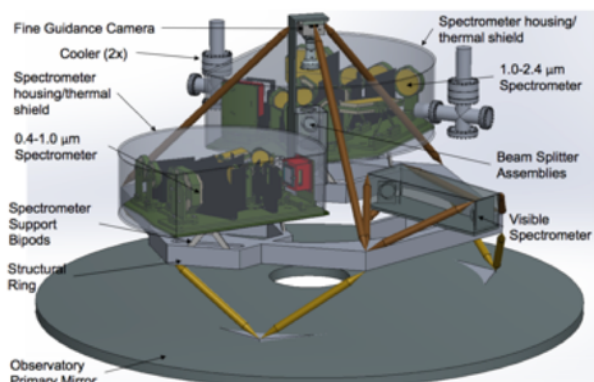

**Figure 2-2**: Schematic design of the EarthFinder instrument suite shows the telescope primary (inverted and at the bottom) and the three spectrometers UVS, OPS, NIRS along with the fine guidance camera.

The payload concept has been developed through a teaming arrangement between JPL, Ball, MIT, George Mason, Notre Dame, and Caltech and is built around a Ball Aerospace spacecraft. The design philosophy is based around ground-based RV instruments, in particular the infrared diffraction limited spectrographs such as PARVI for the Palomar 5 m, iLocator for the Large Binocular Telescope and HISPEC for the Keck Observatory, which use compact very high-resolution cross-dispersed Echelle designs and large format detectors to provide the spectral resolving power and PSF sampling at the detector level to achieve extreme RV precision.

Allocations in RV error budgets were used to flow down into the instrument design. The EarthFinder telescope diameter is driven by the overall photon budget, cost,and the need to complete the primary science mission during the nominal mission lifetime of 5 years.

#### 2.2.3.1   The Telescope

The EarthFinder telescope uses a 1.45 m diameter primary mirror (M1) with an f/# of 1.85, and a total length of 2 m. The secondary mirror (M2) has 0.36 m diameter, resulting in an obscuration ratio of 0.25. The M2 support consists of four horizontal vanes attached to the outer tubular telescope barrel of width about 1 cm each. This is, of course, a notional configuration, which may be modified to ensure that the design and performance meet overall packaging constraints. The core design is a Ritchey-Chretien hyperbola/hyperbola PM/SM arrangement resulting in a total clear aperture of 1.6 m$^2$, with a focal-ratio of 10. Given the warm operation of the telescope (T >225 K), Zerodur is a likely mirror substrate, although silicon carbide is a reasonable alternative at these sizes. Note, however, that the instruments are operated cold, and likely to be made with SiC. A one-time focus mechanism is necessary for the mission and the current baseline is to include a common focus on M3. An alternative will be to have a common focus mechanism within the beam distribution fore-optical switchyard. In addition, EarthFinder provides a small solar aperture (collecting area 2 cm$^2$, FOV 0.36 deg$^2$) that couples sunlight into a 22 cm diam. integrating sphere, which has fiber outputs to the spectrographs.

#### 2.2.3.2   Fine Guidance System

The spacecraft attitude control system is augmented with a payload-provided FGS designed to meet the overall pointing stability needs. The FGS consists of the Fast Guidance Camera (FGC), the Fast Steering Mirror (FSM) and the pointing control algorithms. The EarthFinder FSM design has heritage, and is derived from a high-performance dither mechanism developed earlier for the Space Interferometry Mission (SIM). The FGC uses a deep depletion (with high red sensitivity to 950 nm), low noise CCD capable of windowed frame rates as high as 2 kHz, along with housing and electronics will be based on JPL developed Angular Momentum Desaturation Angle Tracker electronics. The FGC fore-optics reformats the beam to image (Nyquist sampled at 500 nm) the target for efficient guiding (41 mas/pixel plate scale). Starlight for the FGC is previously split off in the EarthFinder fore-optical switchyard. In addition to the main on-axis acquisition and guiding functions, the FGC provides ~1 imaging field of view (FOV) with a 10" calibrated capability for off-axis guiding for possible ancillary science needs.

Each fiber-fed PRV spectrograph is equipped with the capability to back-illuminate from within the instrument SM-fiber bundle to create reference spots on the FGC. The FGC then guides with respect to these reference artificial stars rather than a fixed location on the detector. The optical split between the FGC and the optical spectrograph is determined such that EarthFinder is able to reach the 2-axis guiding noise equivalent angle to stars as faint as V=15 (color temp 5800 K) at 100 Hz rates. For fainter targets, pointing and fiber coupling efficiency degrade with brightness, however coupling into the long wavelength NIRS is nevertheless good for V=18 targets.

### 2.2.3.3  Beamsplitter Optical Switchyard

The primary science needs and observing efficiency requirements dictate that all three spectrographs operate simultaneously. As standard fibers have a limited wavelength range of single mode operation, the RV instruments OPS and NIRS use two independent red-blue input fibers each to receive starlight, resulting in a total of four fibers that need to be fed at the same time (For OPS, possible short and long wavelength fiber arms are in the ranges 400-650 nm and 650-960 nm; for NIRS, these fiber arms can span between 950-1500 nm and 1500-2500 nm). The UVS is slit-fed, limited in total wavelength span and does not require a red-blue split arrangement. Simultaneous delivery of starlight into the respective fibers requires a beam-splitter dichroic arrangement as illustrated in the systems block diagram in **Figure 2-3**. Starting with a pupil image located at the payload FGC, the first split (UV/Optical Infrared) has a beam-splitter transmitting ultraviolet light into the UV spec. slit. A second dichroic split has coatings that partially reflect optical light to the FGC, while reflecting the bulk of the optical-NIR light downstream.

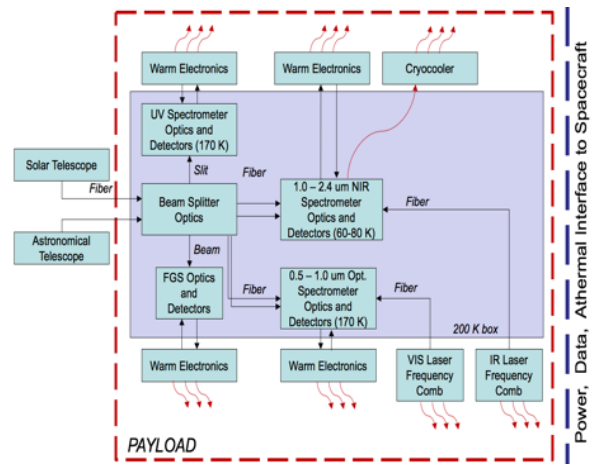

**Figure 2-3:** Payload block diagram showing the feeds three spectrographs and the fast guiding camera.

The last split is a hot mirror arrangement reflecting near infrared light while transmitting the optical beam. Finally, dichroics split optical light the two abovementioned contiguous red-blue bands, delivered into separate silica fibers with ~2-4 µm core diameters. Similarly, the NIR light is split two ways, 950-1500 nm and 1500-2500 nm. The red IR IR arm is delivered to NIRS using fluoride-glass fiber with an ~ 8 µm diameter core. All blue IR port use standard silica fiber. For both spectrometers, a retroreflector arrangement is used to back-illuminate the fibers onto the FGC.

### 2.2.3.4    UV Spectrograph

The operating wavelength span of UVS is 280-380 nm. It has two modes called Band 1 and Band 2. Band 1 covers the entire span at a medium resolving power of R=3000 with 2-pixel sampling per spectral channel and a plate scale of 0.1 nm. The design considered is an Offner in a Littrow configuration, and is implemented with all-spherical optical components. Band 2 spans a narrower wavelength range covering 278-281 nm, and provides a 10x higher resolving power of 30,000 (0.0093 nm per channel). The design strengths are an all-reflective design, minimizing reflective surfaces with high overall transmission. The instrument is useable either the Band 1 or Band 2 mode, with Band 1 being the default mode for RV work.

The UVS detector is a low noise 2kx4.6k back-illuminated E2V CCD42-90, that is delta-doped for ultraviolet QE enhancement.

### 2.2.3.5    Optical and IR Echelle Spectrographs:

The RV spectrographs provide a relatively high optical high efficiency, which EarthFinder trades against the aperture size. Ground-based fiber-fed RV instruments (e.g. High Accuracy Radial Velocity Planet Searcher (HARPS), High Accuracy Radial Velocity Planet Searcher for the Northern Hemisphere (HARPS-N), and NEID) typically suffer a minimum ~50% coupling losses into the spectrometer fiber. This loss is generally due to practical constraints on spectrometer size, and the lack of availability of large-format diffraction gratings. The flux loss, largely dictated by the desire to achieve high resolving power (R~100,000) in seeing-limited operation, is compounded with the transmission losses due to the Earth's atmosphere.

By avoiding these classically-limiting sources of loss, and leveraging precision small optics that can be heavily optimized to maximize efficiency and wavefront, EarthFinder enjoys a significantly higher total system transmission than next generation PRV instruments on the ground (e.g. NEID, EXPRES, and ESPRESSO). Furthermore, the wavelengths spanned by EarthFinder provides significantly more recorded information content per exposure than other current or planned facilities (factor of several beyond NEID, ~8 times that of HARPS), further improving the photon noise floor for a single velocity measurement (**Figure 2-8**).

The OPS and NIRS spectrometers share a common design based on a cross-dispersed white light Echelle to achieve high spectral resolution, $R \sim 170,000 \times (\lambda_0/\lambda)$, where $\lambda_0$=0.6 $\mu$m or 1.6 $\mu$m. **Figure 2-4** gives a schematic of the NIR spectrometer based on a two-arm design for the Keck HISPEC instrument, wherein both short and long wavelength arms use a common Echelle grating but separate fibers. A fiber bundle from the OTA containing the red and blue wavelengths from star and sky as well as the laser frequency comb is fed into the instrument. Collimated light with a 35 mm beam is sent to an R4 Echelle, whose output goes through a beam splitter to either the blue or red wavelength arm, where it is cross-dispersed by either a second grating or prism to be imaged onto a H4RG detector with 10 $\mu$m pitch, and either 1.7 $\mu$m or 2.5 $\mu$m cutoff. The short and long wavelength detectors are cooled to 60-70 K to minimize detector dark current. The Teledyne infrared detectors are mature for spaceflight. H2RG devices are being used in James Webb Space Telescope (JWST) and Euclid while H4RG devices are in advanced development for WFIRST. The detectors will be read out using System for Image Digitization, Enhancement, Control, and Retrieval (SIDECAR) or equivalent cold electronics. A schematic of the broad, cross-dispersed footprint of the echellogram as it appears on the pair of detectors is shown in **Figure 2-5**.

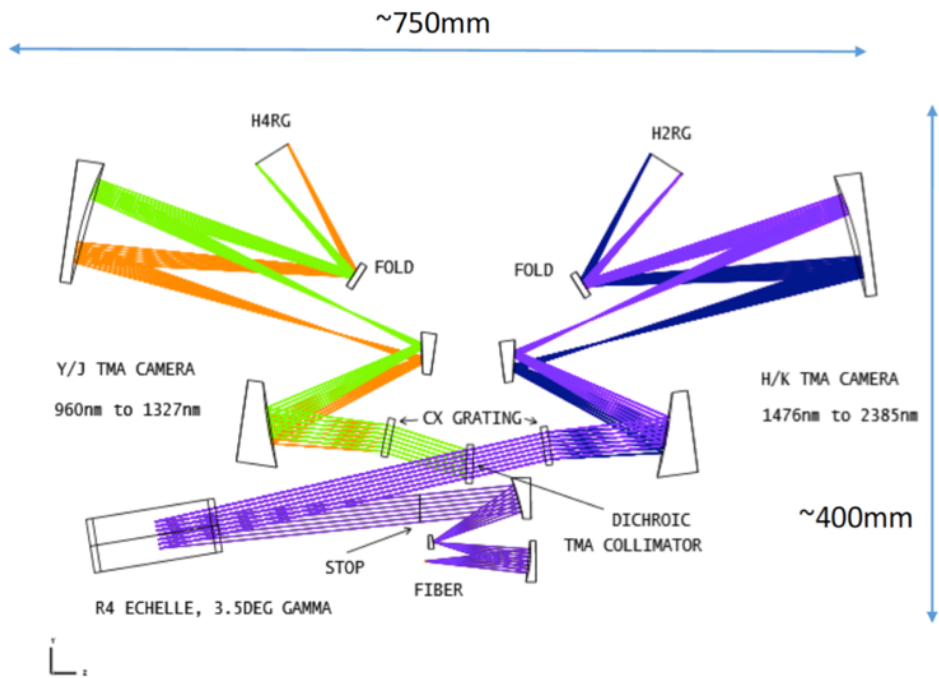

**Figure 2-4:** Schematic layout of a NIRS-like spectrometer based on an instrument concept for the Keck HISPEC instrument. Light enters via a bundle of single mode fibers (short and long wavelengths, sky, optical fiber combs) and is collimated into a ~25 mm diameter beam and projected onto an R4 echelle grating (~10 lines/mm) using a Three Mirror Anastigmat (TMA) . The echelle is slighted tilted to send the outgoing beam to a beamsplitter which sends the light to long and short arms with crossed-dispersing gratings (or prisms) and cameras illuminating Teledyne H4RG/H2RG infrared detectors. The overall scale of the optics footprint is 0.4 m ✕ 0.7 m. Credit: J. Fucik, Caltech Optical Observatories.

The optical spectrograph can follow a similar design recipe, but uses a pair of 9k×9k back-illuminated E2V CCD290-99 detectors with 10 μm pitch. Gratings with diffraction limited performance will be more of a challenge for OPS. **Figure 2-6** shows an illustrative color-coded error budget for a high-resolution Echelles, with error terms listed for an EarthFinder Echelle. As noted in the caption, EarthFinder eliminates or mitigates many of the terms in the error budget by virtue of operating a compact, diffraction-limited instrument fed by a single mode fiber which is monitored via a laser frequency comb angle metrology.

More detailed velocity error-budgets are shown in **Table 2-2** and **Table 2-3**. Our designs which are based on a preliminary thermal assessment carried out by BASD (Section 3), suggest that we can achieve single measurement instrumental stability at the 5 cms$^{-1}$ level through careful thermal design, which holds the spectrograph sandwich stable at the 10 mK level. EarthFinder's instrumental errors will be dominated by second order, non-ideal detector effects such as image memory, non-linearity induced PSF changes, variable interpixel capacitance, pixel response characteristics (QE, higher moments, and pixel locations) in the near IR detectors, and charge transfer inefficiency, bulk thermal changes, and pixel response characteristics and pixel inhomogeneity in CCDs.

Photon noise and stellar jitter must be added separately to the instrument velocity noise (**Figure 2-8**). With the collecting area of a 1.45 m telescope, the time required to achieve a few cm s$^{-1}$ photon-limited precision at the instrument level from the bright stars (median V~5) that direct imaging missions will observe (e.g. the HabEx target list) is ~1 hr.

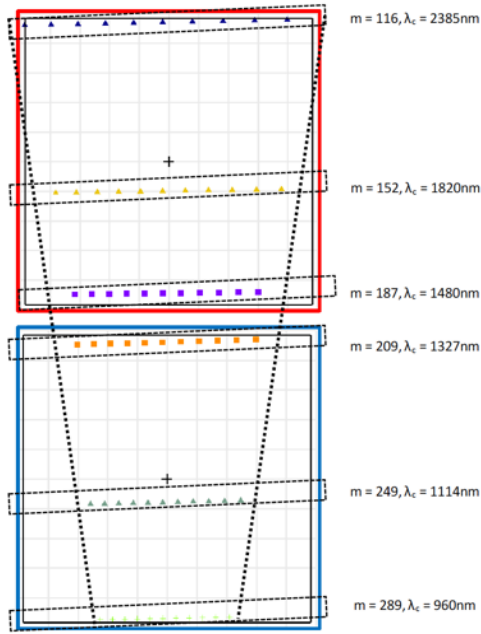

**Figure 2-5**: The Keck HISPEC Echellogram layout using an H2RG and an H4RG to cover 960-2500 nm range. EarthFinder will use two H4RGs. Credit: J. Fucik, Caltech Optical Observatories.

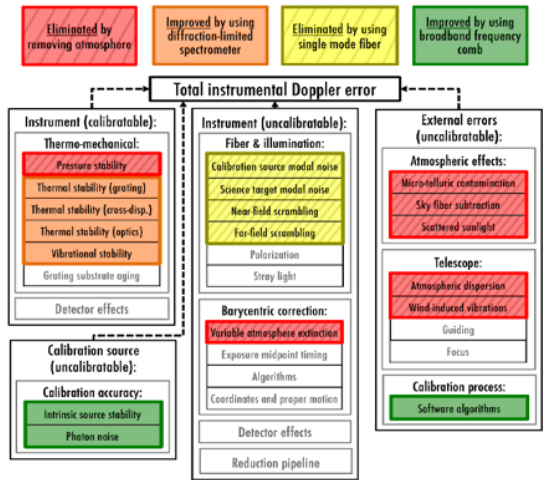

**Figure 2-6**: A generic error budget for a high-resolution RV spectrometer has many terms which the EarthFinder eliminates or mitigates through operation in space (red), use of a diffraction limited spectrometer (orange), and a single-mode fiber (yellow), or via calibration with a laser frequency comb (green). A representative EarthFinder error budget is shown in **Table 2-2** and **Table 2-3**.

The importance of EarthFinder's unmatched wavelength span with high spectral resolving power, and its high cadence observational capability for measuring and mitigating the RV noise from stellar jitter are previously discussed at length in **Section 1.4**.

## 2.3    WAVELENGTH CALIBRATION

The recent ESS report highlights the importance of precision wavelength standards for precision RV measurements for exoplanet research as noted in the finding, "Radial velocity measurements are currently limited by variations in the stellar photosphere, instrumental stability and calibration, and spectral contamination from telluric lines…". (ESS, 2018, pS-1-3)

While efforts to develop the high precision calibration sources needed for ground-based RV studies have led to significant advances in wavelength standards, space-based implementations impose a set of addition requirements; not only must these sources provide an instrument capability at or below 1 cm s$^{-1}$ RV precision, but they must also be compact, low power, and long-lived. Here, we discuss the wavelength calibration options considered for the EarthFinder mission.

### 2.3.1    SPECTROGRAPH CALIBRATION OPTIONS

For decades, the calibration sources used for spectrographs in ground-based observatories have been hollow cathode lamps (typically Thorium-Argon) and gas absorption cells (Iodine) in the visible. Such sources preclude Doppler precision of better than about 0.2 m s$^{-1}$ (Fischer et al. 2016).

A Fabry-Perot (FP) etalon is an alternative calibration source that produces broadband optical combs with the desired density of features when illuminated with a broadband white light source. These etalons offer a low power, relatively compact wavelength solution. Schwab et al. (2015) have engineered etalons locked to the D2 hyperfine transition of rubidium to deliver ~cm s$^{-1}$ - level long-term stability. As such, line referenced etalons are being constructed as calibrator for the Magellan Advanced Radial velocity Observer of Neighboring eXoplanets (MAROON-X), Hermes, Keck Planet Finder (KPF), Fibre-fed Echelle Spectrograph

FIES and iLocater spectrographs (Schwab et al, 2018). FP etalons require precise thermal stabilization to correct for dispersion effects to ensure that stabilizing one fringe is sufficient to stabilize the entire spectrum. They are sensitive to polarization and alignment of input light, and mirror coating degradation. It is also difficult to maintain single mode operation over broad spectral bands.

LFC calibration sources offer the highest possible RV instrument precision achievable today. While the ultimate precision of self-referenced LFCs (< cm/s) may seem to be more than necessary for any individual PRV measurement, it is their long-term precision and stability, compared with that afforded by etalons or arc lamps, that is crucial for achieving the detection of Earth-analogs orbiting solar analogs as called for in the ESS study (ESS, 2018). Thus, there is strong impetus for continued development of laser frequency combs that are suitable as frequency standards for astronomical instruments in space.

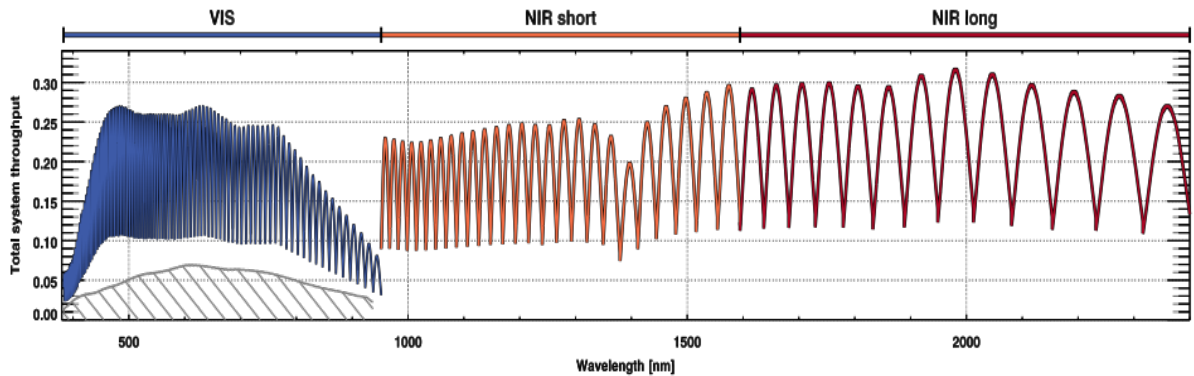

**Figure 2-7:** System throughputs of the OPS and NIRS spectrographs compared with the throughput of the new, next generation NN-EXPLORE NEID spectrograph. The high optical throughput of space-based systems allows for a robust trade with the telescope aperture-diameter.

**Table 2-2**: EarthFinder NIRS PRV Error Budget.

| Instrumental Error (Calibratable) | 16.2 |
|---|---|

| Instrumental Error (Uncalibratable) | 5.2 | | Calibratable Error Contribution | 1.3 | | Calibration Source (Uncalibratable) | 1.5 | | External Error (Uncalibratable) | 1.0 |
|---|---|---|---|---|---|---|---|---|---|---|

**Instrumental Error (Uncalibratable) — 5.2**

| Fiber & Illumination | 2.2 |
|---|---|
| Calibration source modal noise | 0 |
| Continuum modal noise | 0 |
| Near-field scrambling | 0 |
| Far-field scrambling | 0 |
| Stray light and ghosts | 2 |
| Polarization | 1 |
| Focal ratio degradation (science) | 0 |
| Focal ratio degradation (calibration) | 0 |
| Double scrambler mechanical drift | 0 |
| Fiber contamination | 0 |
| Reformater drift | 0 |

| Detector effects | 4.4 |
|---|---|
| Latent images | 3 |
| Pixel inhomogeneity/ Non-linearity | 3 |
| Interpixel capacitance | 1 |

| Barycenter correction | 1.4 |
|---|---|
| Algorithms | 1 |
| Exposure midpoint time | 1 |
| PSF variation | 0 |
| Coordinates and proper motion | 0 |

| Reduction pipeline | 1.0 |
|---|---|
| Software algorithms | 1 |

**Calibratable Error Contribution — 1.3**

| Thermal-Mechanical | 12.3 |
|---|---|
| Thermal stability (grating) | 5 |
| Thermal stability (cross-disperser) | 5 |
| Thermal stability (bench) | 1 |
| Thermal stability (camera) | 5 |
| Optical elements (tilt) | 5 |
| Vibrational stability | 5 |
| Pressure stability | 0 |
| Zerodur phase change | 5 |
| Optical elements (focus) | 1 |

| Detector effects | 10.5 |
|---|---|
| Pixel inhomogeneity | 10 |
| Electronic noise | 1 |
| Pixel location error | 0 |
| Detector thermal expansion | 3 |
| Readout themal thermal transients | 1 |
| CTE | 0 |

**Calibration Source (Uncalibratable) — 1.5**

| Calibration Source (uncalibratable) | 1.1 |
|---|---|
| Wavelength stability | 0.5 |
| Photon noise | 1 |

| Calibration process | 1.0 |
|---|---|
| Software algorithms | 1 |

**External Error (Uncalibratable) — 1.0**

| Telescope/Other | 1.0 |
|---|---|
| Guiding errors | 0 |
| Atmospheric Dispersion Corr. | 0 |
| Telluric | 0 |
| Zodi/Sun/Moon | 1 |

| General Parameter | Value |
|---|---|
| Calibration factor | 0.92 |
| On sky fiber diameter (") | 0.2 |
| Instr. Resolution | 150000 |
| # of science slices | 1 |

| Total Instrumental Error (cm/s) | 5.7 |
|---|---|

**Table 2-3**: EarthFinder OPS PRV Error Budget

| Instrumental Error (Calibratable) | 19.4 |
| --- | --- |

| Instrumental Error (Uncalibratable) | 4.4 |
| --- | --- |

| Fiber & Illumination | 2.2 |
| --- | --- |
| Calibration source modal noise | 0 |
| Continuum modal noise | 0 |
| Near-field scrambling | 0 |
| Far-field scrambling | 0 |
| Stray light and ghosts | 2 |
| Polarization | 1 |
| Focal ratio degradation (science) | 0 |
| Focal ratio degradation (calibration) | 0 |
| Double scrambler mechanical drift | 0 |
| Fiber contamination | 0 |
| Reformater drift | 0 |

| Detector effects | 3.3 |
| --- | --- |
| CTE | 3 |
| Thermal changes | 1 |
| Pixel inhomogeneity | 1 |

| Barycenter correction | 1.4 |
| --- | --- |
| Algorithms | 1 |
| Exposure midpoint time | 1 |
| PSF variation | 0 |
| Coordinates and proper motion | 0 |

| Reduction pipeline | 1.0 |
| --- | --- |
| Software algorithms | 1 |

| Calibratable Error Contribution | 1.5 |
| --- | --- |

| Thermal-Mechanical | 16.3 |
| --- | --- |
| Thermal stability (grating) | 8 |
| Thermal stability (cross-disperser) | 6 |
| Thermal stability (bench) | 1.5 |
| Thermal stability (camera) | 5 |
| Optical elements (tilt) | 7 |
| Vibrational stability | 7 |
| Pressure stability | 0 |
| Zerodur phase change | 6 |
| Optical elements (focus) | 2 |

| Detector effects | 10.5 |
| --- | --- |
| Pixel inhomogeneity | 8 |
| Electronic noise | 1 |
| Pixel location error | 2 |
| Detector thermal expansion | 4 |
| Readout themal thermal transients | 4 |
| CTE | 3 |

| Calibration Source (Uncalibratable) | 1.5 |
| --- | --- |

| Calibration Source (uncalibratable) | 1.1 |
| --- | --- |
| Wavelength stability | 0.5 |
| Photon noise | 1 |

| Calibration process | 1.0 |
| --- | --- |
| Software algorithms | 1 |

| External Error (Uncalibratable) | 1.5 |
| --- | --- |

| Telescope/Other | 1.5 |
| --- | --- |
| Guiding errors | 0 |
| Atmospheric Dispersion Corr. | 0 |
| Telluric | 0 |
| Zodi/Sun/Moon | 1.5 |

| General Parameter | Value |
| --- | --- |
| Calibration factor | 0.92 |
| On sky fiber diameter (") | 0.2 |
| Instr. Resolution | 150000 |
| # of science slices | 1 |

| Total Instrumental Error (cm/s) | 5.1 |
| --- | --- |

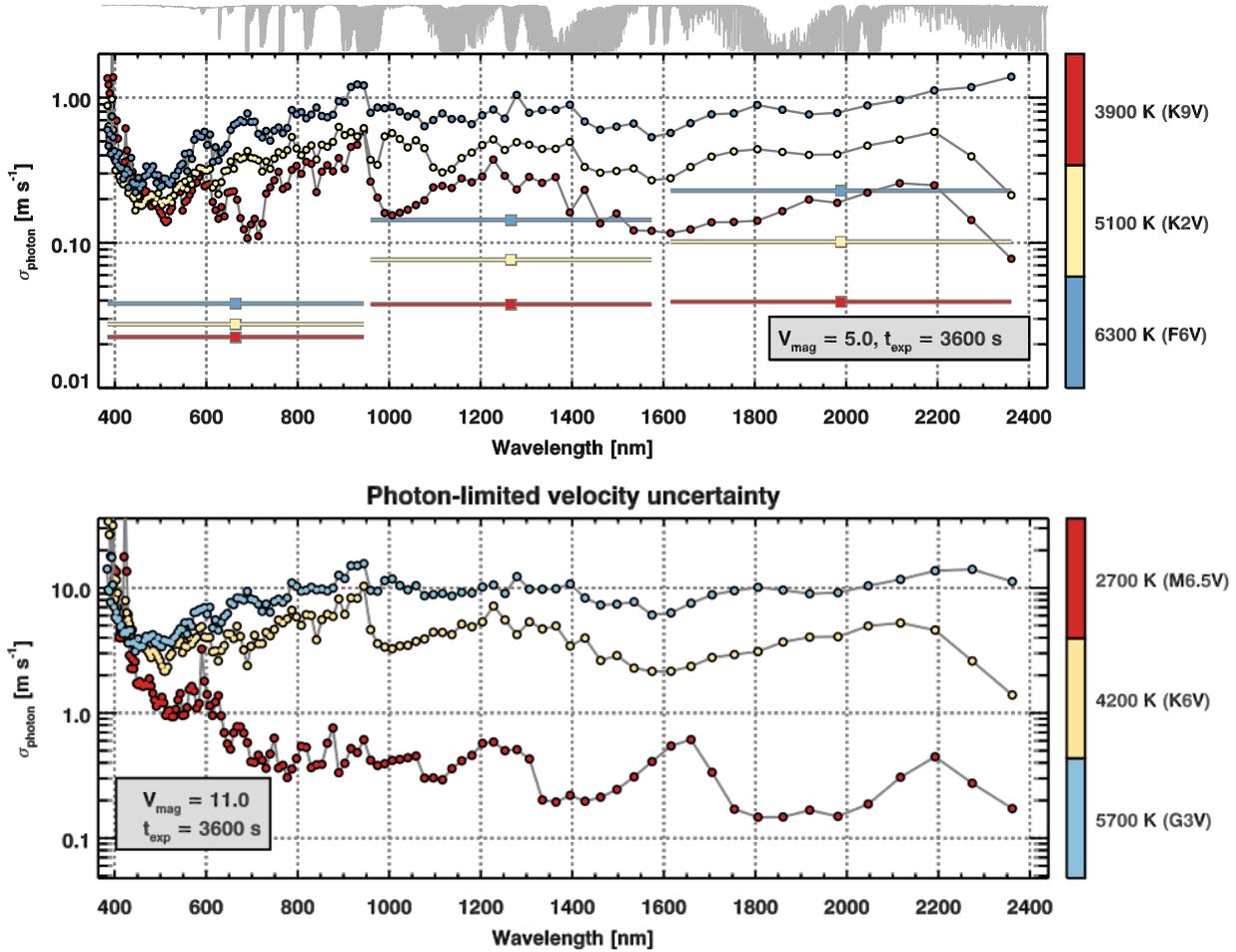

**Figure 2-8**: The EarthFinder photon limited RV precision per spectral order for V=5 and 11 targets (top/bottom panels) with three different effective temperatures, in an hour of integration. The colored points show the achievable precision within individual OPS and NIRS diffraction orders. With EarthFinder's effective area, high measurement precision (<50 cm/s) can be achieved within individual echelle orders on bright targets. Most RV information content for FGK stars is contained in OPS data, while the NIRS channels provide ~ 10 cm/s precision, but crucially chromatic diagnostics for evaluating stellar noise. NIRS provides as much or more information content as OPS when observing the cooler stars.

### 2.3.2    OPTICAL FREQUENCY COMBS AS SPECTRAL RULERS - ASTROCOMBS

One of the earliest and most successful applications of optical frequency combs has been their use as calibration sources for ground-based astronomical spectrographs. The high stability and wide bandwidth of these "spectral rulers" provides the ideal wavelength reference for stellar spectra. While visible band frequency combs for astronomy, a.k.a. astrocombs, were initially based on fiber laser comb technology, the intrinsic free spectral range of these instruments, 100s of MHz to 1 GHz, is too fine to be resolved by astronomical spectrographs of R~150,000 or less. Thus, mode filtering of comb lines to create a more spectrally sparse calibration grid is necessary in these systems. The filtering step introduces complexity and additional sources of instability to the calibration process, as well as instrument assemblies too large in mass and volume for flight. Alternatively, frequency combs produced by electro optic modulation (EOM) of a laser source have been demonstrated at observatories for PRV studies in the near-IR (Halverson et al. 2014 and Yi et al. 2016). EOM combs produce modes spaced at a RF modulation frequency, typically 10-30 GHz, and are inherently suitable as ground-based astrocombs. Significantly, EOM combs avoid the line filtering step of commercial mode-locked fiber laser combs. Comb frequency stabilization can be accomplished in a variety of ways, including referencing the laser pump source to a molecular absorption feature (Yi et al. 2016) or another frequency comb (Halverson et al. 2014). Where octave spanning EOM combs are available, f-2f self-referencing provides the greatest stability (Beha et al. 2015 and Carlson et al. 2017).

EOM combs must be spectrally broadened to provide the octave bandwidth necessary for f-2f stabilization for stability traceable to the SI second. This is accomplished through pulse amplification followed by injection into Highly Non-Linear Fiber (HNLF) or nonlinear optical waveguides, but the broadening process is accompanied by multiplication of the optical phase noise from the EOM comb modulation signal and must be optically filtered. Also, at these challenging microwave pulse repetition rates, the pulse duty-cycle requires pulse amplification to 4-5 Watts of average optical power in order to generate the high enough peak intensity needed for nonlinear broadening. This necessitates use of high power, non-telecom amplifiers that are more prone to lifetime issues, making EOM combs not optimal for flight either. It is interesting to note that very little comb light is actually required on the spectrograph detectors for calibration. In fact, most of the generated comb light must be deliberately attenuated to avoid detector saturation.

In the JPL TeamX study of the EarthFinder mission concept, the power consumption of the frequency comb calibration system was shown to be a significant driver of mission cost, and motivates the development of a comb system that operates with less than 20 Watts of spacecraft power. Thus, for flight applications, it is highly desirable to develop frequency comb technology with low power consumption, ~10 GHz mode spacing, compact size, broad (octave spanning) spectral grasp across both the visible and NIR, phase noise insensitivity, stability traceable to the definition of the SI second, and very importantly, long life.

### 2.3.3  OPTIONS FOR FLIGHT ASTROCOMBS

Within the last year, a very promising path to flight for astrocomb technology possessing the above attributes has been demonstrated by two groups (Obrzud et al. 2018 and Suh et al. 2018). In both efforts, astrocombs were generated through a combination of parametric oscillation and four-wave-mixing in chip-scale ultra-high-Q whispering gallery mode (WGM) optical resonators or microcombs. Microcomb technology has been of keen interest for miniaturization of applications of conventional table-top frequency combs to frequency metrology, time keeping and telecommunications (Diddams 2010). While these microcomb devices have been under study for over a decade, it has only been recently that a significant breakthrough has enabled extremely stable comb formation through the generation of soliton mode locking (Herr et al. 2014, Yi et al. 2015, Brasch et al. 2016, Wang et al. 2016, and Joshi et al. 2016). These solitons, like optical solitons studied in fiber, maintain a stable waveform through a balance of dispersion with the Kerr nonlinearity (Bao et al. 2014). However, in contrast to conventional solitons, the new solitons (referred to as dissipative Kerr solitons or DKSs (Herr et al. 2015)) also have the ability to regenerate using parametric gain that also results from the Kerr nonlinearity. In effect, these regenerating solitons create a mode-locked optical parametric oscillator that functions as a frequency comb. The resulting comb spectral envelopes are found to be very stable and repeatable, which is advantageous for astrocombs. Also, the compact size of the microcombs means that their natural comb line spacing is quite large. In fact, and in contrast to conventional mode-locked combs, microcombs operate more easily with line spacings that are very large (typically 10s of GHz to THz rates).

Suh et al (2018) reported on astronomical spectrograph calibrations with a 22.1 GHz silica soliton microcomb (see **Figure 2-9** showing the 3 mm diameter, quality factor 300 million packaged and pig-tailed resonator) at the Keck Observatory. An image of the soliton comb lines projected onto the NIRSPEC echelle spectrometer is shown in **Figure 2-10**. Suh's soliton microcomb spanned a few tens of nanometers, and was externally spectrally broadened to attain several hundred nanometers of coverage.

A similar approach was demonstrated in a laboratory setting by Lamb et al. (2018) in which a full octave was achieved with a 15 GHz microcomb. Around the same time, Obrzud et al (2018) demonstrated a 23.7 GHz soliton microcomb fabricated from SiN on the GIANO-

B high-resolution near infrared spectrometer at the Telecopio Nazionale Galileo in Spain.

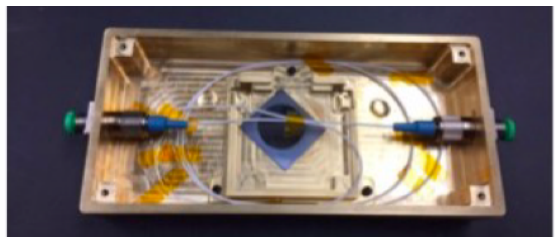

**Figure 2-9:** Silica microastrocomb package demonstrated at the Keck Observatory (Suh et al, 2018).

This comb spectrum spanned a bandwidth similar to Suh's broadened comb, but without an additional broadening stage; instead, the comb was pulse-pumped using a high peak power pump to overcome its much higher resonator loss (i.e., much lower quality factor resonator design). Significantly, both comb designs fell well short of the octave-span goal required for f-2f self-referenced operation. If, however, pulse-pumping in combination with ultra-high quality (UHQ) factor microresonators is pursued, octave-spanning micro-astrocombs at GHz repetition rates may be possible. Pulsed semiconductor laser systems are now capable of producing stable pulses that would be suitable for direct pumping of the octave comb (Nürnberg et al. 2018).

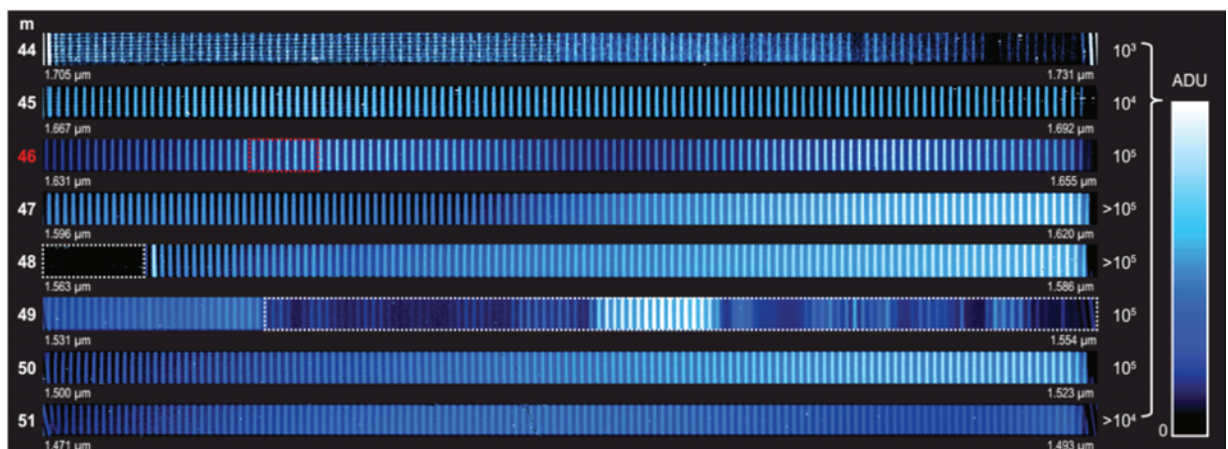

**Figure 2-10:** Image of soliton comb projected onto the NIRSPEC Echelle spectrometer at the Keck Observatory in orders 44 to 51 with the corresponding wavelength ranges of each order indicated. The white dashed box indicates soliton emission and has been heavily filtered to prevent potential damage to the spectrograph detector (Suh et al, 2018). ADU: Analog-to-Digital Units

Despite their low operational power, soliton microcombs will generate comb line power levels that are still 7 orders-of-magnitude larger (10s of microwatts per line) than what is required for spectral calibration in space, thereby retaining the immense available loss margin.

### 2.3.4 FLIGHT COMB REQUIREMENTS AND ARCHITECTURE

To achieve the 380 nm through 2500 nm spectral coverage required for EarthFinder's optical and NIR arms, multiple small form factor comb sources may be necessary. Recent progress has shown a push toward visible band soliton microcomb formation (Lee et al. 2017 and Moille et al. 2018). Other options for EarthFinder's optical arm include second harmonic generation (SHG) of NIR comb lines made in a nonlinear crystal or a Fabry-Perot etalon referenced to a line of the NIR comb in the red portion of the spectrum for stabilization.

Several other subsystem components are necessary for EarthFinder's calibration system, including the pump laser, thermoelectric coolers (TEC), photodetectors for stabilization control circuits, controllers and associated electronics, spectral flatteners to provide uniform power per comb line across the full observation band, and an onboard atomic clock (such as is pictured in **Figure 2-11**) in a GPS-denied environment for RF stabilization of the comb repetition frequency.

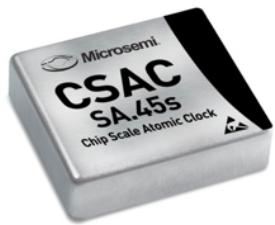

**Figure 2-11:** The Chip Scale Atomic Clock (CSAC): A RF Frequency Reference for the comb repetition rate stabilization will be needed where no GPS signal is available.

**Table 2-4** shows the power consumption of components used in a compact, self-referenced, low (~250 MHz) repetition rate fiber laser comb. Many of the elements in this system would also be needed in a soliton microcomb-based flight astrocomb system.

**Table 2-5** shows various pulse-pumped microcomb concepts considered during the JPL TeamX study for EarthFinder. The concepts reflect decreasing required power levels for lower TRL technology. At the highest TRL are EOM-pumped microcombs; these would consume too much (>100 W) power for a two-comb system. Some advances to the EOM stages (TRL 2) would result in about 60 W of required power for the combined subsystem. Pulsed semiconductor laser pumped systems or dual microcomb systems (microcomb-pumped microcombs) are the lowest TRL while offering the highest potential mass, power, and volume savings. This last technique of dual microcombs is inspired by the recent demonstration of a two-comb frequency synthesizer developed under DARPA sponsorship (Spencer et al, 2017).

**Table 2-4:** Power consumption of a compact fiber laser comb package, housing consisting of one 18 cm x 20 cm x 2.5 cm box (Sinclair et al, 2015). Many of the elements in this system would also be used in a soliton microcomb-based flight comb system.

| Comb System Element | Typical Current (mA) | Voltage (V) | Typical Power (W) |
|---|---|---|---|
| 1480 nm diode laser | 700 | 2.7 | 2(x2) |
| 980 nm diode laser | 1500 | 2.7 | 4(x2) |
| Diode laser (TEC) | 500 | 2.7 | 1(x4) |
| PPLN TEC | -- | -- | <1 |
| Femtosecond laser housing TECs | 1000 | 8 | 8 |
| Commercial photodetectors for $f_{opt}$, $f_{ceo}$, and $f_{rep}$ | <250 | 15 | <4(x3) |
| FPGA-based controller[a] | 1800 | 5 | 9 |
| Control electronics[b] | 200 | | 5 |
| **Total for comb system** | **--** | **--** | **50** |

## 2.3.5 PATH FORWARD – TECHNOLOGY ROADMAP

Of all the technologies considered for an EarthFinder mission, microcomb-based calibration standards are the least mature. Nevertheless, pulse-pumped microresonator technology is advancing rapidly under support from agencies like DARPA for applications to time standards and frequency synthesis. Moreover, these miniature combs satisfy key requirements for astronomy such as 5-30 GHz native mode spacing and a technology platform to support operation outside the controlled laboratory environment. However, further advances are necessary to extend this technology to preserve the mid-repetition rate regime (i.e., higher than fiber laser combs and lower than Terahertz microcombs) while simultaneously achieving octave-spanning spectral grasp without the addition of a nonlinear optical broadening stage.

A technology roadmap for EarthFinder's calibration sources would include the following elements:

1. Extend silica and possibly Si3N4 microcombs to create pulse-pumped microastrocombs capable of delivering octave spanning spectra at ~10 GHz repetition rates in the soliton regime. Requires soliton microcomb dispersion engineering to maintain coupling efficiency of pumping wave and allow broad comb formation with pulsed excitation.

a. Demonstration of small form-factor pulse pumping source (e.g. microcomb pumped by pulsed semiconductor laser)

b. Packaging: incorporate fully integrated waveguide structures with the comb microresonator.

2. Continued rubidium D2 line-locked FP etalon development through exploration of advanced material designs. Improved thermal stability of these devices will help overcome spectral effects typical of all etalons.

**Table 2-5:** Advanced microcomb calibration system power estimates and components for 4 comb architecture concepts.

| Component | Component TRL | EOM-pumped Soliton Microcomb TRL3 | EOM-pumped Soliton Microcomb TRL2 * requires specially engineered modulator | Pulsed Semiconductor Laser-pumped Soliton Microcomb TRL 2 | Dual Soliton Comb TRL1 |
|---|---|---|---|---|---|
| Microcomb resonator | 2-3 | 0 | 0 | 0 | 0 |
| CW pump laser (DFB) | 6 | 4 | 4 | - | 4 |
| Laser TEC | 6 | 1 | 1 | 1 | 1 |
| PPL drive oscillator | 6 | 8 | 8 | 8 | - |
| CSAC | 6 | 0.12 | 0.12 | 0.12 | 0.12 |
| RF amplifier | 5 | 36 | 10 | - | - |
| Modulators | 2-4 | 0 | 0 | | |
| EDFA 1 | 5 | 4 | 4 | - | - |
| EDFA 2 | 5 | 2.5 | 2.5 | 2.5 | 2.5 |
| Pulse semiconductor laser | 3 | - | - | 4 | - |
| Misc. electrically actuated attenuators, switches, control electronics, photodetectors | 4-5 | 4 | 4 | 4 | 4 |
| **Total per comb** | | **58** | **30** | **20** | **10** |
| **Total system** | | **116 W** | **60 W** | **40 W** | **20 W** |

**Table 2-6**: Payload Mass and Power

| Instrument | Component Characteristics and Mass | Power (W) |
| --- | --- | --- |
| | CBE Mass | Power Mode 2 |
| | kg | Operating |
| **NICM Costing Total** | **197.4** | **357.5** |
| Detector | 3.8 | ^Includes 43% contingency |
| Electronics | 38.5 | |
| Optics | 67.8 | |
| Thermal | 38.0 | |
| Structure | 49.4 | |
| NIR Spectrometer (1.0 µm-2.5 µm) | 51.2 | 6.0 |
| VIS Spectrometer (0.5 µm-1.0 µm) | 40.6 | 14.0 |
| UV Spectrometer | 8.5 | 10.0 |
| Fine Guidance | 11.0 | 0.0 |
| Frequency Laser Comb | 30.0 | 60.0 |
| Beam Splitter | 2.3 | 0.0 |
| Cryocoolers | 25.5 | 600.0 |
| Assembly-Level Structure/Thermal | 62.5 | 160.0 |
| **Instrument Total** | **231.51** | **850** |

# 3    SPECTROGRAPH THERMAL EVALUATION

Spectrograph stability is paramount for achieving EarthFinder's precision RVs; any small perturbations within the spectrograph(s) optical train, either global optic positions or surface figures, will lead to motion of the incident beam on the focal plane. In the mission spectrograph models, a $1/10000^{th}$ of a pixel shift is a velocity error of 6 cms$^{-1}$. These perturbations can arise from numerous sources, however, variations in the thermal environment leading to changes in the opto-mechanics is regarded as a key driver of overall instrument precision. While LFC calibration is used to correct for this effect, there will always be some residual calibration error and therefore to achieve precisions at the few cms$^{-1}$ level, current ground-based instruments have demonstrated the need for, and adopted, routine instrument thermal stability at the mK levels within their vacuum chambers.

To develop an initial design and cost estimate for EarthFinder, it has been important to assess and understand the thermal requirements for the instrument spectrographs. This was carried out via an engineering study undertaken by Ball Aerospace, which focused on utilizing designs and specifications from diffraction limited spectrographs with similar design characteristics to EarthFinder. The design of the iLocater spectrograph (currently under development at the University of Notre Dame for the Large Binocular Telescope, AZ, USA) was principally adopted for this work as it offers some of the close similarities to a future EarthFinder design: it is single-mode fed, operates in the NIR (0.97-1.31 um), uses and H4RG detector from Teledyne, and is designed to achieve high resolving power (R>150,000; Crepp et al. 2016) akin to EarthFinder. The instrument can be illuminated simultaneous by three single-mode optical fibers to generate three spectral traces per diffraction order. These can be used for calibration, sky and science light. For the study it was assumed that a single main on-axis fiber would be used for a science delivery and one or both of the remaining fibers would inject calibration light.

The STOP analysis focused on determining the required thermal-optical stability limits needed to achieve the RV precision defined by the EarthFinder science cases. Using an optical and CAD model of iLocater, finite element simulations were undertaken to assess relative motion between PSFs of the three input fibers when imaged at the detector plane. Any relative change between the centroids of these PSFs will generate an error between the calibration source and star. Therefore, by requiring this error to have minimal impact on science performance, a limit can be imposed on the thermal stability requirements for any design. The study calculated these deviations for nine monochromatic wavelengths of light distributed across the instrument detector.

CODE V was used to determine the sensitivity of each optic within the system (6 degrees of freedom) and its impact on PSF location at the instrument focal plane. The sensitivity analysis itself does not itself provide requirements on stability; it simply assess the impact on PSF position in the instrument focal plane when an optic is perturbed in a specific axis with a specific magnitude. This information has to be combined with how the optics themselves move under changing thermal conditions.

The sensitivity study analysis was combined with a finite element model of the mechanical instrument. This model was used to determine the exact motions of each optic with changing temperature. Together, this allows the impact of thermal changes to be realistically assessed at the instrument focal plane. For this analysis, the instrument was assumed to be fabricated from a single uniform SiC base material, which, while not fully representative, significantly simplifies the simulation process. It allows reasonable bounding estimates of the thermal requirements to be generated within the time and budget constraints of this study.

Changes in the position of any optic or optomechanical assembly are primarily driven by material used for fabrication and its coefficient of thermal expansion (CTE). The combined finite element analyses and optical sensitivities were combined and the allowable thermal change was assessed against the design limit of 0.27 nm of motion in dispersion direction at the detector focal plane. The initial iLocater design assumed an all-aluminum system providing a thermal stability requirement of 1-2 mK, consistent with previous analysis completed as part of the development of iLocater. Aluminum, however, does not offer an optimal CTE value for stability, and therefore SiC was chosen to drive the instrument thermal requirements.

As the CTE of a material is often temperature dependent, the instantaneous CTE at the instrument operating temperature must be used when studying thermal stability requirements. By carefully choosing an operating temperature, is it often possible to utilize an optimal value of a material CTE, further loosening any thermal stability requirements. Silicon Carbide at an operating temperature of 80K (instantaneous CTE = $0.16 \times 10^{-6}$) was assumed, from which the study derived a thermal stability requirement of 60 mK.

The results of the analysis show significant benefits to utilizing an intrinsically stable SiC or Zerodur sandwich bench, and tuning the instrument temperature to achieve the optimal CTE value during operation. This allows a loosening of the thermal stability requirements for the instrument. The analysis completed herein had several assumptions (isotropic and homogeneous materials), which are optimistic for a real instrument. We, therefore, adopted a more cautious approach and defined the thermal stability requirements of ±10mK long term for both spectrographs, OPS and NIRS. This assumes using optimized materials for each operating temperature and implementing more precise thermal control on individual components (e.g. detectors) as needed.

# 4    COST, RISK, HERITAGE ASSESSMENT

## 4.1    COST ASSESSMENT

The primary goal of the NASA EarthFinder Probe study was to investigate the necessity of a space mission to achieve the PRV precision required to detect Earth analogs orbiting the nearest bright stars which would also be the targets for future direct imaging missions. Thus, no high-fidelity instrument or mission design studies were funded. However, JPL was able to carry out a TeamX study to establish whether EarthFinder was consistent with the anticipated cost of a Probe Class mission.

The TeamX cost estimate of $755M ($905M including launch vehicle in FY18 dollars) is presented in **Table 4-1**. The estimate includes 30% of unreserved costs as cost reserves as required by JPL best practices. The TeamX estimate is based on a detailed estimate of the (WBS 5) payload system costs, and rule of thumb percentages for the other WBS elements of the mission. The TeamX detailed payload system estimate is based the NASA Instrument Cost Model (NICM) version VIII for the Instrument (Fine Guidance Camera, all three spectrometer arms, the beam splitter, and the laser comb); a multivariable parametric cost model by Stahl & Henrichs (2016) for the Optical Telescope Assembly (OTA); and the NICM VIII cryocooler cost model for the 60K Cryocooler.

There is one significant difference between the version of EarthFinder studied by TeamX and the version developed later in the study is the telescope aperture. TeamX assumed a telescope diameter of 1.1 m whereas the current version discussed elsewhere in this report is 1.45 m. There was no funding to iterate with TeamX after the larger telescope aperture was adopted. A new trade study will be required between OTA cost ($\propto D^{1.4\sim2}$; Stahl and Hendrichs 2016) and performance, e.g. number of stars, number of observations, single measurement precision.

**Disclaimer:** The costs presented in this report are ROM estimates; they are not point estimates or cost commitments. It is possible that each estimate could range from as much as 20% percent higher to 10% lower. The costs presented are based on Pre-Phase A design information, which is subject to change. The cost information contained in this document is of a budgetary and planning nature and is intended for informational purposes only. It does not constitute a commitment on the part of JPL and/or Caltech.

**Table 4-1**: Cost Estimate

| Work Breakdown Structure (WBS) Elements | Team X Estimate |
|---|---|
| 1.0, 2.0, & 3.0 Management, Systems Engineering, and Mission Assurance | $58.22 M |
| 4.0 Science | $15.65 M |
| 5.0 Payload System | $263.02M |
| 5.01 Payload Mgmt. | - |
| 5.02 Payload SE | - |
| 5.03 Payload S&MA | $4.75 M |
| 5.04 OTA | $44.10 M |
| 5.05 Instrument | $193.58 M |
| 5.01 Inst. Mgmt. | $12.19 M |
| 5.02 Inst. SE | $13.32 M |
| 5.03 Inst. S&MA | $7.04 M |
| 5.04 Sensor | $151.01 M |
| 5.03 60K Cryocooler | $10.03 M |
| 5.10 Instrument I&T | $20.59 M |
| 6.0 Flight System | $183.59 M |
| 7.0 & 9.0 Mission Op Preparation & Ground Data Systems | $30.08 M |
| 10.0 ATLO | $23.57 M |
| 11.0 Education and Public Outreach | - |
| 12.0 Mission and Navigation Design | $6.75 M |
| Reserves (30%) | $174.26 M |
| 8.0 Launch Vehicle (LV) | $150.00 M |
| **Total Cost (including LV)** | **$905.14 M** |

# A. ACRONYMS

| AGI | Analytical Graphics, Inc. |
|---|---|
| AMD | Angular Momentum Desaturation |
| CBE | Current Best Estimate |
| CCD | Charge-coupled Device |
| CFE | Combined Finite Element |
| CMOS | Complementary Metal-oxide Semiconductor |
| CODE V | Optics Design Program |
| CTE | Coefficient of Thermal Expansion |
| DARPA | Defense Advanced Research Projects Agency |
| EMCCD | Electron-multiplying Charge-coupled Device |
| EOM | Electro Optic Modulation |
| ESPRESSO | Echelle Spectrograph for Rocky Exoplanet and Stable Spectroscopy Observations |
| FGC | Fine Guidance Camera |
| FGS | Fine Guidance System |
| FIES | Fibre-fed Echelle Spectrograph |
| FOV | Field of View |
| FP | Fabry-Perot |
| FSM | Fine Steering Mirror |
| FWHM | Full Width Half Maximum |
| GHz | Gigahertz |
| GIANO-B | |
| GPS | Global Positioning System |
| HabEx | Habitable Exoplanet Imaging Mission |
| HARPS | High Accuracy Radial Velocity Planet Searcher |
| HARPS-N | High Accuracy Radial Velocity Planet Searcher for the Northern Hemisphere |
| HISPEC | High Resolution Infrared Spectrograph |
| HNLF | Highly Non-Linear Fiber |
| HST | Hubble Space Telescope |
| IR | Infrared |
| IRSPEC | Infrared Spectrometer |
| JWST | James Webb Space Telescope |
| KPF | Keck Planet Finder |
| LFC | Laser Frequency Comb |
| LSF | Line Spread Function |
| MAROON-X | Magellan Advanced Radial velocity Observer of Neighboring eXoplanets |
| MIT | Massachusetts Institute of Technology |
| MLI | Multi-layer Insulation |
| NAS | National Academy of Sciences |
| NEID | NN-EXPLORE Exoplanet Investigations with Doppler Spectroscopy |
| NIRS | Near-Infrared Spectroscopy |
| NIRSPEC | Near-Infrared Spectrograph |
| NRS | Non-rotationally Symmetric |
| OIR | Optical Infrared |
| OPS | Open Path Spectrometer |
| OTA | Optical Telescope Assembly |
| PARVI | Palomar Habitable Zone Planet Finder |
| PM | Primary Mirror |

| PRV | Precise Radial Velocity |
|---|---|
| PSF | Point Spread Function |
| QE | Quantum Efficiency |
| RC | Ritchey–Chrétien |
| RMS | Root Mean Square |
| RV | Radial Velocity |
| S/C | Spacecraft |
| SiC | Silicon Carbide |
| SIDECAR | System for Image Digitization, Enhancement, Control, and Retrieval |
| SIM | Space Interferometry Mission |
| SM | Single Mode |
| SM | Secondary Mirror |
| STM | Science Traceability Matrix |
| TEC | Thermoelectric Coolers |
| UHQ | Ultra-high Quality |
| UVS | Ultraviolet Spectrograph |
| WFIRST | Wide Field Infrared Survey Telescope |
| WGM | Whispering Gallery Modes |